\documentclass[sigconf, nonacm]{acmart}

\usepackage{pvldb}

\renewcommand\vldbdoi{XX.XX/XXX.XX}
\renewcommand\vldbpages{XXX-XXX}
\renewcommand\vldbavailabilityurl{URL_TO_YOUR_ARTIFACTS}

\usepackage{xspace}
\usepackage{amsmath,amsfonts}
\usepackage{stfloats}
\usepackage{graphicx}

\theoremstyle{acmdefinition}
\newtheorem{example}{Example}
\theoremstyle{acmplain}

\usepackage[ruled, noline, noend, linesnumbered]{algorithm2e}

\SetAlFnt{\small}
\SetAlCapFnt{\small}
\SetAlCapNameFnt{\small}

\usepackage{subcaption}
\usepackage{booktabs}
\usepackage{enumitem}
\usepackage{cleveref}

\newcommand{\btitle}[1]{\noindent\textbf{#1}}
\newcommand{\refalg}[1]{Algorithm~\ref{alg:#1}}
\newcommand{\refsec}[1]{Section~\ref{sec:#1}}
\newcommand{\reftab}[1]{Table~\ref{tab:#1}}

\newcommand{\kw}[1]{{\ensuremath {\mathsf{#1}}}\xspace}
\newcommand{\algpr}{\kw{PageRank\text{-}Dynamic}}
\newcommand{\algprDelta}{\kw{Delta\text{-}PageRank\text{-}Dynamic}}
\newcommand{\algcc}{\kw{CC\text{-}Dynamic}}
\newcommand{\algsssp}{\kw{SSSP\text{-}Dynamic}}
\newcommand{\algmis}{\kw{MIS\text{-}Dynamic}}
\newcommand{\algcolor}{\kw{Color\text{-}Dynamic}}

\newcommand{\apicompute}{\kw{Compute}}
\newcommand{\apiexplore}{\kw{Explore}}
\newcommand{\apisend}{\kw{send}}
\newcommand{\apiactivate}{\kw{activate}}
\newcommand{\apifilter}{\kw{Filter}}
\newcommand{\apigather}{\kw{Gather}}
\newcommand{\apifornb}{\textbf{for}~nb \in nbr}
\newcommand{\apiforins}{\textbf{for}~nb \in ins}
\newcommand{\apifordel}{\textbf{for}~nb \in del}
\newcommand{\apivertex}{\kw{vertex}}
\newcommand{\apiloadupdate}{\kw{LoadUpdate}}
\newcommand{\apideployupdate}{\kw{DeployUpdate}}

\SetKwProg{fn}{function}{$:$}{}
\SetKwProg{stat}{statement}{:}{}
\newcommand{\kwbool}{\kw{bool}}
\newcommand{\kwint}{\kw{int}}
\newcommand{\kwlong}{\kw{long}}

\begin{document}
\title{A Flexible and High-Performance Framework for Distributed Dynamic Graph Processing}

\author{Kaiyu Chen}
\affiliation{\institution{The University of New South Wales}
  \city{Sydney}
  \country{Australia}
}
\email{kaiyu.chen1@unsw.edu.au}

\author{Junhua Zhang}
\affiliation{\institution{Northeastern University}
  \city{Shenyang}
  \country{China}
}
\email{zhangjunhua@neu.edu.cn}

\author{Dong Wen}
\affiliation{\institution{The University of New South Wales}
  \city{Sydney}
  \country{Australia}
}
\email{dong.wen@unsw.edu.au}

\author{Lantian Xu}
\affiliation{\institution{University of Technology Sydney}
  \city{Sydney}
  \country{Australia}
}
\email{lantian.xu@student.uts.edu.au}

\author{Wentao Li}
\affiliation{\institution{University of Leicester}
  \city{Leicester}
  \country{United Kingdom}
}
\email{livent@126.com}

\author{Long Yuan}
\affiliation{\institution{Wuhan University of Technology}
  \city{Wuhan}
  \country{China}
}
\email{longyuan@whut.edu.cn}

\author{Xubo Wang}
\affiliation{\institution{University of Technology Sydney}
  \city{Sydney}
  \country{Australia}
}
\email{xubo.wang@uts.edu.au}

\author{Ying Zhang}
\affiliation{\institution{Zhejiang Gongshang University}
  \city{Hangzhou}
  \country{China}
}
\email{ying.zhang@zjgsu.edu.cn}

\author{Lu Qin}
\affiliation{\institution{University of Technology Sydney}
  \city{Sydney}
  \country{Australia}
}
\email{lu.qin@uts.edu.au}

\begin{abstract}
The proliferation of massive, constantly evolving graphs in domains such as social and financial networks necessitates efficient and scalable dynamic graph processing. Existing distributed systems, such as Ingress, attempt to automatically adapt static algorithms to dynamic updates, but they face critical limitations: overly conservative strategy selection, inability to leverage algorithm-specific optimizations, and limited inference capabilities that often force fallbacks to slow, generic methods.
To overcome these limitations, we propose a new paradigm that emphasizes user flexibility and performance over unreliable automation.
Following this paradigm, we present a high-performance, user-friendly distributed system for dynamic graph processing. Built on the Bulk Synchronous Parallel (BSP) model, it provides a flexible programming model based on vertex subsets to control computation flow, along with a simple API offering powerful abstractions. By enabling direct expression of dynamic logic and incorporating system-level optimizations, our system achieves significant performance gains. We demonstrate its expressiveness through six implemented dynamic graph algorithms, and extensive experiments on eight real-world datasets validate its superior performance and flexibility compared to state-of-the-art solutions.
\end{abstract}

\maketitle

\vldbtopmatter

\section{Introduction}\label{sec:introduction}
Real-world graphs, such as social networks and web graphs, are typically massive in scale and highly dynamic~\cite{garg2009evolution,ntoulas2004s}. The dynamic nature of these large graphs necessitates scalable and efficient processing capabilities to extract both structural and temporal insights. For instance, in financial fraud detection or e-commerce analytics, dynamic graph analysis is crucial for identifying suspicious users and preventing fraudulent transactions~\cite{cen2019trust,jiang2016understanding}.

To support these applications, researchers have explored \emph{Incremental Graph Processing}~\cite{solomon2020improved,DBLP:journals/pvldb/YuanQLCZ17,lin2021hierarchical,karczmarz2024fully,stix2004finding}—a paradigm that updates prior results in response to graph changes. However, real-world graphs are large in scale, and designing and implementing distributed algorithms for these graphs is notoriously complex~\cite{doekemeijer2014survey}. Hence, significant work has been done on building distributed graph processing systems
~\cite{DBLP:conf/sigmod/MalewiczABDHLC10,DBLP:conf/osdi/GonzalezXDCFS14,low2012distributed,DBLP:conf/osdi/ZhuCZM16,DBLP:conf/icde/LiM0LYQ0Z23,DBLP:conf/www/YanCLN15} to provide users with easy-to-use, high-performance environments for implementing distributed graph algorithms. However, these systems are primarily designed for static graph algorithms and cannot support dynamic graph algorithms.

\btitle{Existing Solutions.}
Recently, a distributed system, Ingress~\cite{DBLP:journals/pvldb/GongTYYZGYYZ21,DBLP:journals/vldb/GongTYWYZYGFYZ24}, that focuses on incremental graph processing has emerged.
Ingress is an automated system for incremental graph processing that aims to simplify the implementation of dynamic graph algorithms.
Ingress supports a spectrum of iterative graph algorithms that can be expressed in a vertex-centric manner.
Ingress accepts a static graph algorithm with three functions: \emph{Aggregation}, \emph{Update} and \emph{Propagation} for aggregating messages from neighbors, updating vertex values based on the incoming messages and generating messages to neighbors, respectively.
Given the three functions, Ingress automatically analyzes the properties of these functions, such as associativity, commutativity, etc., using Satisfiability Modulo Theories (SMT) solvers~\cite{DBLP:conf/tacas/MouraB08}.
Based on the analyzing results, Ingress automatically selects a best-fit strategy among \emph{Memorize Free}, \emph{Memorize Path}, \emph{Memorize Vertex} and \emph{Memorize All} to support efficient incremental execution of the provided static graph algorithm.
The \emph{Memorize Free} strategy stores nothing for the incremental processing and generate compensation and cancellation messages to offset the effect of invalid messages upon graph updates.
While the \emph{Memorize Path} strategy traces the dependency tree of vertex values, it resets and recomputes the values of vertices in the subtree disconnected from a deleted edge in the dependency tree.
The \emph{Memorize Vertex} strategy records the values of each vertex in all iterations, and recomputes the values of vertices that rely on the graph update.
The \emph{Memorize All} strategy stores all messages and vertex values in all iterations are stored in the system, it recomputes the messages and vertex values that rely on the graph update.
This strategy is the fallbacks when none of the other three strategies can be applied.

\btitle{Limitations.}
Although the automatic incrementalization of static graph algorithms is convenient for users and demonstrates the effectiveness of the memorization strategies in supporting incremental execution, it has several drawbacks that limit its applicability to a wide range of high-performance applications.

\emph{\underline{Overly Conservative Strategy Selection.}}
The conditions for selecting efficient memorization strategies are often overly conservative. As a result, many algorithms revert to a slow, generic default strategy that stores all messages exchanged during computation—even when a lightweight memorization strategy would suffice—causing severe performance degradation.
Consider the $k$-Core Decomposition~\cite{dorogovtsev2006k}, which computes the core number of each vertex, defined as the largest integer $k$ such that the vertex belongs to a $k$-core (a maximal subgraph in which every vertex has degree at least $k$). A vertex-centric iterative algorithm initializes each vertex's core number to its degree; in every iteration, each vertex broadcasts its current core number to its neighbors and updates its own value using the H-index of the received numbers~\cite{lu2016h}. This process continues until all vertices converge. Because the core number of a vertex at iteration $i$ depends only on its neighbors' values at iteration $i-1$, an incremental strategy is to store the core number of each vertex across iterations and recompute only those affected by updates. This corresponds to the \emph{memorize vertex} strategy in Ingress.
However, Ingress requires that the update function satisfy strict conditions (Condition C1 in Section 4.1~\cite{DBLP:journals/pvldb/GongTYYZGYYZ21}, which demands an inverse function to cancel the effect of invalid messages). The H-index does not meet this requirement, forcing Ingress to fall back to the \emph{memorize all} strategy, which introduces substantial overhead.

\begin{figure}
    \centering
    \includegraphics[width=\linewidth]{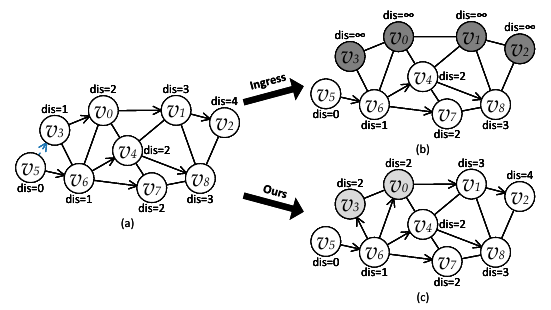}
    \caption{Incremental Processing of SSSP ($v_5$ is the source).}
    \label{fig:sssp}
\vspace{-1.5em}
\end{figure}

\emph{\underline{Incapable of Algorithm-Specific Optimizations.}}
Ingress adopts a one-size-fits-all framework for incremental graph processing, which makes it simple and easy to use. However, this generality comes at the cost of losing opportunities for algorithm-specific optimizations. For instance, the \emph{Memorize Path} strategy in Ingress resets all vertices in the subtree disconnected from a deleted edge in the dependency tree before recomputing their values. While this conservative approach ensures correctness for arbitrary algorithms, it is unnecessarily costly for certain cases.
\Cref{fig:sssp} illustrates this issue with the Single-Source Shortest Path (SSSP) algorithm. The source vertex is $v_5$, and \Cref{fig:sssp}(a) shows the shortest path tree rooted at $v_5$. When the edge $(v_5,v_3)$ is deleted, Ingress resets the distances of $v_3$, $v_0$, $v_1$, and $v_2$ to infinity and recomputes them on the updated graph, as shown in \Cref{fig:sssp}(b). However, this reset is excessive: $v_0$, $v_1$, and $v_2$ still have valid shortest paths through $v_6$ with the same distances as before. A more efficient update strategy, shown in \Cref{fig:sssp}(c), is to recompute only the distance of $v_3$ while leaving $v_0$, $v_1$, and $v_2$ unchanged, except for updating the predecessor of $v_0$ to $v_6$.

\emph{\underline{Limited Inference Capabilities.}}
The automation mechanism in Ingress introduces another source of unreliability. Ingress relies on SMT solvers (e.g., Z3~\cite{DBLP:conf/tacas/MouraB08}) to analyze user-defined functions and verify whether they satisfy certain properties. However, as noted in Section~5.2 of Ingress~\cite{DBLP:journals/pvldb/GongTYYZGYYZ21}, these solvers have limited inference power and may fail to prove a property even when it actually holds. In such cases, the system must conservatively assume the worst case and revert to a safe but inefficient default strategy.

\btitle{Our Solution.}
Realizing the limitations of existing systems and the challenges of building an automated framework that is both high-performance and general, we take a different approach: empowering users to directly express their dynamic graph algorithms within an easy-to-use, high-performance distributed system. Leveraging the proven simplicity and effectiveness of the Bulk Synchronous Parallel (BSP) model~\cite{DBLP:conf/sigmod/MalewiczABDHLC10} for distributed graph processing, our system is designed around BSP. We provide a flexible API that enables users to access and manipulate graph updates, allowing them to express dynamic graph algorithms in a simple and intuitive manner. We name our system \emph{\underline{VersaGraph}}—short for \underline{Versa}tile \underline{Graph} Processing System—to emphasize its strengths in flexibility, performance, productivity, and user-friendliness for dynamic graph processing. In particular, VersaGraph is characterized by the following key designs and features:

\emph{\underline{Flexible Programming Model.}}
Unlike static graph algorithms, which can often be expressed as simple iterative functions, dynamic graph algorithms are more complex and require sophisticated control flow and operators.
To address this, our system API is based on the concept of a \emph{Vertex Subset}, which models active vertices similarly to the Pregel framework~\cite{DBLP:conf/sigmod/MalewiczABDHLC10}.
Users can modify the vertex subset and control when computations are performed on it, enabling fine-grained control over the algorithm's execution flow. In addition, users have direct access to graph updates.
To simplify update handling, we provide two interfaces, $\apiloadupdate$ and $\apideployupdate$, for loading and deploying updates, respectively. Together, these features allow our system to support a wide range of dynamic graph algorithms, surpassing the capabilities of existing systems.

\emph{\underline{High Performance.}}
Our system achieves high performance through two approaches: (1) enabling users to implement algorithms in a way optimized for their specific characteristics, and (2) optimizing the system itself. In addition to the flexible programming model, we provide two primary computation paradigms, \emph{Push} and \emph{Pull}, whose performance depends on the particular algorithm and graph structure~\cite{besta2017push}, allowing users to choose the most suitable approach.
Beyond these algorithm-level optimizations, the system is carefully engineered for high performance, incorporating in-memory edge caching, communication latency hiding, and synchronization restricted to critical attributes. These combined measures allow VersaGraph to efficiently handle dynamic graph workloads while maintaining flexibility and expressiveness.

\emph{\underline{High Productivity and User-Friendliness.}}
VersaGraph offers simple, intuitive APIs that let users implement dynamic graph algorithms in just a few lines of code with minimal boilerplate. The API follows a functional style and revolves around two core abstractions, \apicompute and \apiexplore, covering most algorithm logic.
With the notion of statements for directly manipulating vertex states and incoming messages, users can focus entirely on the core algorithm, while the system handles execution details. Even complex dynamic graph algorithms can be expressed concisely, demonstrating VersaGraph's ease of use, high productivity, and user-friendliness.

\btitle{Contributions.}
In summary, our contributions are as follows:
\begin{itemize}[leftmargin=2\labelsep]
    \item \emph{Analysis of Existing Systems.}
    We systematically analyze current dynamic graph processing systems and identify their key limitations, including overly conservative strategy selection, lack of support for algorithm-specific optimizations, and limited inference capabilities.
    \item \emph{A New Paradigm for Dynamic Graph Processing.}
    We propose a new paradigm that prioritizes user flexibility over automation, enabling users to directly express dynamic graph algorithms with high performance and expressiveness.
    \item \emph{API and System Design.}
    Guided by this paradigm, we develop a flexible programming model and intuitive APIs characterized by their expressiveness and simplicity in implementing dynamic graph algorithms. Additionally, the system is also carefully optimized for high performance through multiple system-level enhancements.
    \item \emph{Extensive Evaluation.}
    We implement six dynamic graph algorithms to showcase the expressiveness and flexibility of our programming model. Extensive experiments on a cluster of eight cloud instances using eight real-world datasets demonstrate that our system outperforms existing solutions.
\end{itemize}

\btitle{Paper Organization.}
The next section presents the basic concepts and problem definition. \Cref{sec:framework} introduces the programming model and API design of our framework, followed by system implementation details and optimizations in \Cref{sec:implementation}. \Cref{sec:app} showcases several dynamic graph algorithms implemented using our framework, and \Cref{sec:exp} evaluates the system's performance. Finally, we review related work in \Cref{sec:related} and conclude the paper in \Cref{sec:conclusion}.
\section{Preliminaries}
\btitle{Graphs.}
Let $G(V,E)$ denote a graph containing a set of vertices $V$ and a set of edges $E$. The numbers of vertices and edges are denoted by $|V|$ and $|E|$, respectively.
For dynamic graphs, the graph structure may change over time due to edge or vertex insertions and deletions. We use $\Delta G$ to denote a set of graph updates that consists of edge insertions $\Delta E_{ins}$, edge deletions $\Delta E_{del}$, vertex insertions $\Delta V_{ins}$, and vertex deletions $\Delta V_{del}$, i.e., $\Delta G = (\Delta E_{ins}, \Delta E_{del}, \Delta V_{ins}, \Delta V_{del})$, and $G \oplus \Delta G$ to denote the updated graph after applying $\Delta G$ to $G$. For simplicity, we assume that there are only edge insertions and deletions in this paper, i.e., $\Delta G = (\Delta E_{ins}, \Delta E_{del})$, as vertex insertions and deletions can be handled by inserting or deleting isolated vertices and inserting or deleting their adjacent edges, where the insertion and deletion of isolated vertices can be trivially handled.

\btitle{Static Graph Algorithms.}
Given a static graph algorithm $\mathcal{S}$, it takes a static graph $G$ as input and produces a result $R=\mathcal{S}(G)$ as output, and it does not change the input graph or rely on dynamic updates to produce the output.

\btitle{Dynamic Graph Algorithms.}
Given a static graph algorithm $\mathcal{S}$, the goal of a dynamic graph algorithm $\mathcal{D}$ for $\mathcal{S}$ is to efficiently maintain the result $R$ of $\mathcal{S}$ after a graph update $\Delta G$. Specifically, $\mathcal{D}$ takes a static graph $G$, the output $R=\mathcal{S}(G)$, and a graph update $\Delta G$ as input, and produces an updated result $R'=\mathcal{D}(G,\Delta G,R)$ as output, such that $R'$ is the result of $\mathcal{S}(G')$, where $G' = G \oplus \Delta G$.

\btitle{Graph Processing Systems.}
Processing large-scale graphs efficiently is challenging due to the complexities in implementing distributed algorithms~\cite{doekemeijer2014survey}. Many systems~\cite{DBLP:conf/sigmod/MalewiczABDHLC10,DBLP:conf/ppopp/ShunB13,DBLP:conf/osdi/GonzalezXDCFS14,DBLP:journals/corr/LowGKBGH14,DBLP:conf/osdi/ZhuCZM16,DBLP:conf/icde/LiM0LYQ0Z23,DBLP:conf/www/YanCLN15} have been proposed to facilitate the development of distributed graph algorithms by abstracting away these complexities. Most of these systems adopt the vertex-centric programming model~\cite{DBLP:conf/sigmod/MalewiczABDHLC10}. They iteratively execute user-defined functions on vertices in parallel. The user-defined functions typically use the data from adjacent vertices or incoming messages to update the local state of the vertex and send messages to neighboring vertices.

\btitle{Problem Statement.}
As most existing distributed graph processing systems are designed for static graphs, the goal of this paper is to design a distributed dynamic graph processing system such that (1) given a graph algorithm $\mathcal{S}$ in the static graph processing system, it is easy to implement a dynamic version $\mathcal{D}$ for $\mathcal{S}$ in the dynamic graph processing system; (2) the dynamic graph algorithm $\mathcal{D}$ is efficient and scalable in the dynamic graph processing system.
\section{The Framework}
\label{sec:framework}
In this section, we present our framework for distributed dynamic graph processing.
The framework adopts a vertex-centric computation model, where each vertex processes its local state and exchanges messages with its neighbors in parallel across supersteps.
For ease of understanding, we first illustrate the framework using a simple PageRank implementation.
We then introduce the details of the programming interface (API) and finally demonstrate the use of combiners and aggregators with an optimized PageRank implementation.

\begin{algorithm}[t]
\caption{$\algpr$}
\KwIn{PageRank \kw{r} of each vertex, graph update $\Delta G$}
\label{alg:pr2}
\tcc{$\apivertex(\kw{double~r})$ --- vertex definition}

$A \gets \apiloadupdate(\Delta G)$\;
$A.\apideployupdate()$\;
$B \gets A.\apicompute(\mathcal{C})$\;
\While{$|B|>0$}{
    $B \gets B.\apiexplore(\mathcal{E}).\apicompute(\mathcal{C})$\;
}
\vspace{0.5em}
\stat{$\mathcal{C}$}{
    $v.\kw{r} \gets 0.15$\;
    $\textbf{for } u\in nbr\textbf{: } \kw{r}~ += 0.85 * \frac{u.\kw{r}}{u.deg}$\;
}
\vspace{0.5em}
\stat{$\mathcal{E}$}{$\textbf{for } u\in v.nbr\textbf{: }\apiactivate(u.id)$\;}
\end{algorithm}

\subsection{Example: A Basic PageRank}
PageRank~\cite{DBLP:journals/cn/BrinP98} is an algorithm that measures the importance of vertices in a graph.
\refalg{pr2} presents the \algpr for computing the PageRank values when a graph is updated.
In this algorithm, each vertex maintains a user-defined attribute $\kw{r}$ representing its PageRank value, and the PageRank values for all vertices have been precomputed for the initial graph.
When a batch of updates arrives, the algorithm begins by loading these updates into the system (line~1) using the \apiloadupdate function, which returns a vertex set $A$ containing all vertices incident to the edges in $\Delta G$.
Subsequently, it deploys the update (line~2) to the system using \apideployupdate to update the graph topology and vertex properties before computation, as the PageRank update process is based on the updated graph.
The algorithm then invokes the \apicompute function on the vertex set $A$ (line~3).
The \apicompute takes a statement $\mathcal{C}$ as input, which defines how to compute the PageRank value for each vertex (lines~6-8).
We use the standard teleportation probability 0.15 and damping factor 0.85 \cite{DBLP:conf/eurosys/MariappanV19, DBLP:conf/eurosys/MariappanCV21}.
The \apicompute function follows synchronous execution semantics, ensuring that PageRank value updates for vertices remain invisible to their neighbors until all computations for vertices in $A$ have finished.
The \apicompute function returns a subset $B$ of $A$, containing vertices whose PageRank values changed beyond a specific threshold $\epsilon$.
As long as $B$ is not empty (line~4), the neighbors of vertices in $B$ are activated by invoking the \apiexplore function on $B$ (line~5), which applies the statement $\mathcal{E}$ to each vertex in $B$ (line~9--10).
The PageRank values for the activated neighbors are then recomputed by applying the \apicompute function on the set of active vertices (lines~5).
This iterative procedure repeats until $B$ becomes empty, indicating that no PageRank value changes exceed the threshold $\epsilon$ and computation has converged.

\subsection{Programming Interface Abstractions}
\subsubsection{Data Model.}

As is demonstrated in the PageRank example, the programming interface of our framework is designed around vertex sets, we refer to it as the \kw{vSet} type in the framework.
Given a (directed) graph $G(V,E)$, our framework maintains several default attributes for each vertex $v$:
\begin{itemize}[leftmargin=2\labelsep]
    \item $id$: the unique identifier of the vertex;
    \item $nbr (nbr^+/nbr^-)$: the set of (in/out) neighbors of the vertex;
    \item $deg (deg^+/deg^-)$: the (in/out) degree of the vertex;
    \item $ins (ins^+/ins^-)$: the set of inserted (out/in) neighbors;
    \item $del (del^+/del^-)$: the set of deleted (out/in) neighbors.
\end{itemize}

Note that the (in/out) degree equals the size of the corresponding neighbor set, i.e., $deg=|nbr| $, $deg^+ = |nbr^+|$ and $deg^- = |nbr^-|$.
Unless otherwise specified, we use the undirected graph $G(V,E)$ as the default graph model.
Apart from the default attributes, each vertex may have user-defined attributes. The PageRank value \kw{r} in the previous section is an example of a user-defined attribute.

\subsubsection{Graph Update Functions.}
Our framework provides two explicit functions to handle graph updates. Rather than modifying the graph directly, updates are staged and deployed in two separate steps.
This allows algorithms to inspect and process updates before committing them. We provide two core functions in particular: $\apiloadupdate$, which loads a batch of updates into temporary attributes, and $\apideployupdate$, which applies the updates to the graph and maintains the consistency of its structural attributes:

\underline{$\apiloadupdate(\Delta G)\rightarrow \kw{vSet}$.}
Each time a batch of graph updates $\Delta G$ arrives, the algorithm is triggered to process the updates.
The main goal of this function $\apiloadupdate(\Delta G)$ is to load the updates into the system and initialize the attributes inserted neighbors $ins$ and deleted neighbors $del$ for each vertex accordingly. During this process, the neighbor attribute $nbr$ of each vertex remains unchanged.
After loading the updates, the function returns a vertex set $A$ containing all vertices incident to the edges in $\Delta G$, i.e., $A = \bigcup_{(u,v)\in \Delta G} \{u,v\}$.
We assume all the batch updates are valid updates, in other words, for any $v\in G$, we have $v.ins\cap v.nbr = \emptyset$ and $v.del\subseteq v.nbr$.
If the batch updates are not valid, the framework will preprocess them so that the updates are valid.

\underline{$\kw{vSet}.\apideployupdate()$.}
This function deploys the updates loaded by $\apiloadupdate(\Delta G)$ to the graph.
This operation is performed on a vertex set, which is typically the vertex set returned by $\apiloadupdate(\Delta G)$.
Specifically, it first applies the changes stored in the attributes $ins$ and $del$ to the neighbors attribute $nbr$ of each vertex, i.e., $nbr = nbr \cup ins \setminus del$, and then updates the degree attributes $deg$ accordingly.
During this process, the attributes $ins$ and $del$ remain unchanged and are not cleared until the next call to $\apiloadupdate(\Delta G)$.

\subsubsection{Statements.}
Before we delve into the graph programming interfaces, we first introduce the notion of \emph{statements}, denoted as \kw{stats} in the framework, which serves as the fundamental building blocks of computation, allowing users to specify the operations to be performed on each vertex in a vertex set.
A statement is syntactically a single or a series of C++ statements with each individual statement separated by a semicolon.
When a statement is passed to a function, it will be enclosed in a lambda function which provides the statement an execution context including the local variables and the data specified by the lambda functions.
For instance, both \apicompute and \apiexplore functions in the framework accept statements as parameters, they enclose the statements in a lambda function, and the execution context, e.g., each vertex, is provided to the statements via parameters of the lambda function.
Statements in the \apicompute function have access to the attributes of a vertex and its neighbors, whereas statements in the \apiexplore function may instead generate messages to the neighbors.
By supporting statements as composable building blocks, the framework achieves goals of flexibility and expressiveness: users can express a wide range of graph operations using a few lines of code.

\subsubsection{Core Graph Programming Interfaces.}
The functions $\apicompute$, $\apiexplore$, $\apigather$, and $\apifilter$ accept $statements$ as input and support various operations:

    \underline{$\kw{vSet}.\apicompute(\kw{stats})\rightarrow \kw{vSet}$.}
The \apicompute function is mainly used for implementing the pull semantics of graph computation (e.g. line~3 in \refalg{pr2}) or processing incoming messages.
This function executes the given statements for each vertex in a vertex set, and returns a set of vertices whose attributes are modified.
During execution, the statements can only read or write the user-defined attributes of each vertex, but they can access the incoming messages and all the attributes of a vertex, including the degree and the user-defined attributes of its neighbors.
The execution of the statements in the \apicompute function is synchronous, which means that even if the attributes of a neighbor are modified before the execution of the statements on the current vertex, these modifications will not be visible until the \apicompute function returns.

\underline{$\kw{vSet}.\apiexplore(\kw{stats}[, \kw{combiner}])\rightarrow \kw{vSet}$.}
The \apiexplore function is mainly used for sending messages to the neighbors. When combined with the \apicompute function, it can implement the push semantics of graph computation.
Similar to \apicompute, this function executes the given statements (e.g., sending messages) for each vertex in a specified vertex set.
It returns a set of vertices that have received messages, which can be processed by the \apicompute function on the returned vertex set.
It can read the attributes of the vertices in $\kw{vSet}$ but cannot change them.
In addition, this function accepts an optional argument \kw{combiner} to aggregate messages with the same target before sending them out.
If the combiner is not specified, the messages sent to the same target vertex will be simply grouped together before sending out.

\underline{$\apiactivate(vid), \apisend(vid, message)$.}
These two functions are used in the statements of the \apiexplore function to send messages to a specific vertex.
Both functions take a vertex identifier $vid$ as input, which specifies the target vertex to receive messages.
We provide two functions here to support different use cases: (1) in some scenarios (line~5 of \Cref{alg:pr2}), we need to perform computations on the neighbors of a vertex without sending any messages, in this case, we can use the \apiactivate function to activate the neighbors without sending any messages; (2) in other scenarios, we need to send data to the neighbors, in this case, we can use the \apisend function to send messages to the neighbors. The underlying implementation of these two functions is the same, we provide two functions with different names in order to support different computation semantics.

\underline{$\kw{vSet}.\apigather(\kw{stats}[,\kw{combiner}])$.}
This function is used to gather data specified in the $statements$ from the vertices in $\kw{vSet}$.
The statements passed to this function must return the data to be gathered, which will be combined using the optional $combiner$ function if provided.
This function is useful for collecting statistical information or attributes from a set of vertices.
For example, we can use $A.\apigather(\text{return } deg, \kw{sum})$ to gather the total degree of vertices in $A$.
The statements passed to \apigather can read all the attributes of the vertices in $\kw{vSet}$.

\underline{$\kw{vSet}.\apifilter(\kw{stats})\rightarrow \kw{vSet}$.}
This function filters vertices in $\kw{vSet}$ based on the given statements.
The statements passed to this function must return a boolean value, which indicates whether the vertex should be included in the filtered set.
The statements have read access to all attributes of a vertex.
For example, if we want to get the set of vertices with the highest degree, we can first use $d_{max}\gets V.\apigather(\text{return } deg, \kw{max})$ to get the maximum degree, and then use $V.\apifilter(\text{return } deg == d_{max})$ get the set of vertices with the highest degree.

\underline{Set Operations.} The \kw{vSet} type supports standard set operations including cardinality $|\kw{vSet}|$, union $A \cup B$, intersection $A \cap B$, and difference $A \setminus B$.

\begin{algorithm}[t]
\caption{$\algprDelta$}
\KwIn{PageRank $\kw{r}$ of each vertex; graph update $\Delta G$}
\label{alg:pr}
\tcc{$\apivertex(\kw{double}~\kw{r},\kw{double}~\kw{dlt})$}

$A \gets \apiloadupdate(\Delta G)$\;

$B \gets A.\apiexplore(\mathcal{E},\kw{sum}).\apicompute(\mathcal{C})$\;

$A.\apideployupdate()$\;

\While{$|B|>0$}{
    $B \gets B.\apiexplore(\mathcal{E}',\kw{sum})$.$\apicompute(\mathcal{C})$\;
}

\vspace{0.5em}
\stat{{$\mathcal{E}$}}{ $d_{new} \gets deg + |ins| - |del|$\;
    $\textbf{for}~u \in ins\!:\apisend(u.id,\frac{\kw{r}}{d_{new}})$\;
    $\textbf{for}~u \in del\!:\apisend(u.id,-\frac{\kw{r}}{deg})$\;
    $\textbf{for}~u \in nbr \setminus del\!:\apisend(u.id,\frac{\kw{r}}{d_{new}}-\frac{\kw{r}}{deg})$\;}
\vspace{0.5em}
\stat{{$\mathcal{C}$}}{$\kw{dlt}\gets 0.85*\kw{sum}(msgs)$; $\kw{r}\mathrel{{+}{=}}\kw{dlt}$\;}
\vspace{0.5em}
\stat{{$\mathcal{E}'$}}{$\textbf{for}~u \in nbr\!:\apisend(u.id,\frac{\kw{dlt}}{deg})$\;}
\end{algorithm}

\subsection{Example: A Delta-Based PageRank}

In this section, we demonstrate the usage of our framework in implementing a Delta-based PageRank algorithm~\cite{gonzalez2012powergraph,zhang2013maiter} using the push computation semantics.
In the Delta-based PageRank algorithm, only when a vertex's PageRank value changes beyond a threshold $\epsilon$, the difference is propagated to its neighbors; as a result, the number of messages sent is significantly reduced compared to the original PageRank algorithm.
Due to the reduced number of active vertices and sparse message propagation, implementing the algorithm in push mode will yield better performance compared to the pull mode.

\Cref{alg:pr} presents the \algprDelta algorithm.
Apart from the PageRank value \kw{r}, each vertex also maintains an additional attribute \kw{dlt} to store the change in PageRank value.
The algorithm first loads the update (lines~1) and then creates and sends the rectifying messages to reflect the neighbor changes of each vertex (lines~2).
Specifically, for inserted neighbors (line~8), a positive correction value is sent to them, while for deleted neighbors (line~9), a negative correction value is sent(line~5). For other neighbors, as the degree of the vertex may have changed, a correction value is also computed and sent to them (line~10).
An aggregation function \kw{sum} is also passed to the \apiexplore to aggregate the correction values sent to the same target vertex (line~2).
These messages are then processed by the \apicompute function (line~2).
Next, the updates are deployed to the graph (line~3).
The algorithm further propagates the changes of PageRank values in the updated graph iteratively until convergence (line~4--5).
\section{Implementation}\label{sec:implementation}

In this section, we introduce the various aspects of the implementation of our distributed dynamic graph processing framework. The framework is designed for in-memory processing of dynamic graphs and imposes minimal constraints on how the graph and its updates are stored or accessed, as long as the graph can be loaded efficiently and updates can be fed into the framework correctly.

\subsection{Graph Partition}
The graph is partitioned using edge-cut methods, in which vertices are grouped into disjoint subsets and each edge is assigned to its incident vertices. Given a cluster with $m$ workers, the graph is divided into $m$ partitions. Each worker manages one partition and stores the master copy of each vertex in that partition. A worker does not necessarily correspond to a physical machine, and multiple workers can run on the same machine to improve resource utilization. If the endpoints of an edge are assigned to different partitions, a mirror copy of each endpoint is created on the worker that does not store the master copy. Both the master copy and its mirrors maintain the attributes of the vertex.

The default partitioning strategy is based on the hash value of vertex IDs, which is simple yet effective for general graphs. The framework also supports custom partitioning strategies. For specific types of graphs, such as web graphs or road networks, placing vertices from the same site or geographic region on the same worker can significantly reduce communication costs and improve performance.

\subsection{The Vertex Subset}
The vertex subset type \kw{vSet} is implemented as a distributed data structure. Each worker maintains the IDs of vertices in its local partition. Given two instances of Vertex Subset $A$ and $B$, their union, intersection, and difference are computed locally on each worker. The cardinality of a vertex subset is obtained by aggregating local sizes from all workers.

Although the provided algorithms are designed for single-batch updates, the framework extends this model to support continuous updates. It monitors incoming changes and processes them at configurable intervals. Updates are buffered until either their count reaches a user-defined threshold or a time limit expires, after which they are dispatched to the corresponding workers according to the partitioning scheme.

\subsection{Graph Update}
Although the algorithms we provide are designed for single-batch updates, the framework supports continuous updates by monitoring graph updates and processing them in a timely manner, with the update interval configurable by users.
Before the framework starts processing updates, it collects incoming updates until either the number of collected updates reaches a specified threshold or a specified time interval elapses. Once either condition is met, the collected updates are routed to the corresponding workers according to the partitioning scheme.

When the framework starts processing updates, it passes the collected updates to the graph algorithm as input. The framework provides two APIs, $\apiloadupdate$ and $\apideployupdate$, to handle graph updates in two stages. In the first stage, each worker invokes $\apiloadupdate$ to store the received updates in two separate attributes for each vertex: $ins$ for inserted neighbors and $del$ for deleted neighbors. Simultaneously, each worker returns a vertex subset containing the affected vertices, i.e., vertices incident to the inserted or deleted edges. In the second stage, $\apideployupdate$ applies the collected updates by adding edges from $ins$ to $nbr$ and removing edges in $del$ from $nbr$, producing updated neighbor lists.

\begin{figure}
    \centering
    \includegraphics[width=\linewidth]{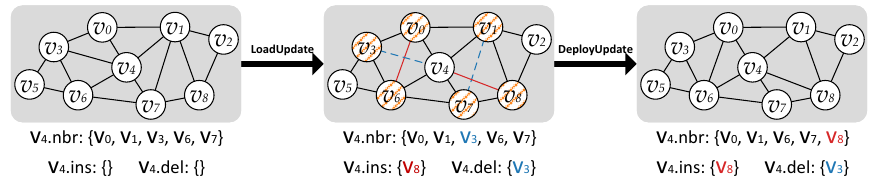}
    \caption{Graph update process.}
    \label{fig:update}
\vspace{-1.5em}
\end{figure}

If a dynamic graph algorithm requires handling updates in a more fine-grained manner, such as processing insertions and deletions separately, users can load and deploy one subset of updates at a time and repeat the process until all updates are processed.

\begin{example}
\Cref{fig:update} illustrates the graph update process. Before the update, the neighbors of $v_4$ are $\{v_0, v_1, v_3, v_6, v_7\}$. After loading the update, red solid edges indicate inserted edges, blue dotted edges indicate deleted edges, and shaded vertices are affected vertices, i.e., those incident to inserted or deleted edges. For example, $v_4.ins = \{v_8\}$ denotes the inserted neighbor of $v_4$, and $v_4.del = \{v_3\}$ denotes the deleted neighbor. After deploying the update, the inserted and deleted neighbors are applied, resulting in the updated neighbor list $v_4.nbr: \{v_0, v_1, v_6, v_7, v_8\}$.
\end{example}

\subsection{Internals of the Core Programming APIs}
Although the graph algorithms in the framework are written in a single-threaded style, they are executed in a distributed manner across multiple workers.
Each worker is a separate instance of the same program, distinguished by a unique worker ID and responsible for processing a different graph partition.
Our framework follows the Bulk Synchronous Parallel (BSP) model, and the computations are divided into distinct supersteps.
Synchronization of the supersteps among workers is achieved implicitly through data exchange or explicitly through barrier synchronization.

\emph{\underline{Implementation of $\apicompute$.}}
The $\apicompute$ plays the most important role among the core APIs and is used to achieve both push and pull computation modes.
We first explain how it works in pull mode.
In pull mode, each vertex reads data from its neighbors to perform computations and modify its attributes.
This process is efficient because each vertex can access its neighbors' attributes locally through their master copies or mirrors.
To guarantee the synchronous computation semantics of $\apicompute$, any modifications to a vertex's attributes are temporarily stored in local memory and remain invisible to other vertices.
This prevents premature access to changes by neighbors.
After the computation finishes, the new state of each vertex is compared with its previous state to check whether the vertex has changed.
If so, the attributes of the master copy are updated and synchronized to its mirrors, ensuring that all vertices see the latest state in the next superstep.
Before $\apicompute$ returns, these changed vertices are collected into a vertex subset and returned to the caller.
In push mode, $\apicompute$ behaves similarly to pull mode, except that it processes incoming messages from neighboring vertices and updates the vertex attributes accordingly.

\emph{\underline{Implementation of $\apiexplore$.}}
The $\apiexplore$ is mainly used to activate or send messages to neighbor vertices in the push mode.
The $\apiactivate$ function is essentially a specialized form of $\apisend$, where an empty message is sent to the target vertex as an activation signal.
Each time $\apisend$ is called, the message is aggregated into the message buffer of the target vertex.
If a combiner is specified, messages sent to the same vertex are combined to reduce communication cost.
Otherwise, they are appended to the message buffer of the target vertex.
After the vertices finish sending messages, the message buffers are sent to the corresponding workers that hold the master copies of the target vertices.
The set of vertices that receive the messages is then returned to the caller.

\emph{\underline{Implementation of $\apigather$ and $\apifilter$.}}
The $\apigather$ function aggregates values over a set of vertices. Each worker maintains a local value to aggregate the values of its local vertices, and these local values are then aggregated to obtain the final result, which is returned to the caller.
The $\apifilter$ function executes the provided statements on each vertex in the subset locally and returns the set of vertices that satisfy the filtering condition.

\begin{figure}
    \centering
    \includegraphics[width=\linewidth]{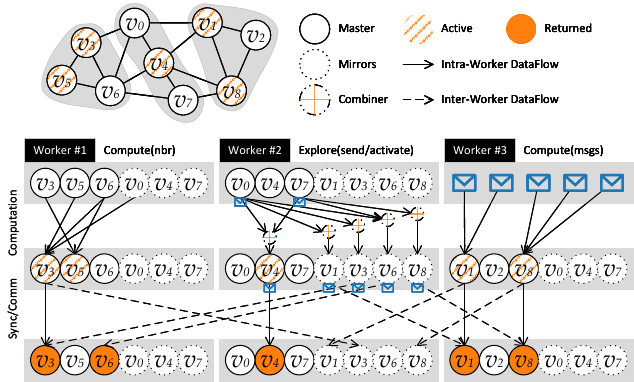}
    \caption{Dataflow of the Core APIs: \apicompute and \apiexplore.}
    \label{fig:arch}
    \vspace{-1.5em}
\end{figure}

\begin{example}
\Cref{fig:arch} illustrates the system data flow for a graph partitioned across three workers.
On Worker~1, vertices $v_3$ and $v_5$ invoke the $\apicompute$ API based on their neighbor lists ($v_3.nbr = \{v_5, v_6, v_0\}$ and $v_5.nbr = \{v_3, v_6\}$). The intra-worker data flow reflects this process. Since $v_3$'s attributes change, it is returned and synchronized to its mirrors. In contrast, $v_5$'s attributes remain unchanged, so it is neither returned nor synchronized.
On Worker~2, vertices $v_0$ and $v_7$ call the $\apiexplore$ API to activate or send messages to their neighbors. Data is first passed in memory to the combiner before communication. For example, $v_0$ and $v_7$ each send a message to $v_6$, which are aggregated by a user-defined function (or packed into a single message) before transmission. Vertex $v_4$ receives messages, becomes active, and is returned. Vertices $v_1$, $v_3$, $v_6$, and $v_8$ also receive messages and are returned on their respective workers.
On Worker~3, after a vertex receives messages, it calls the $\apicompute$ API to process them. If its attributes change, it is returned and the updated values are synchronized to its mirrors, as shown in the sync/comm stage of the figure.
\end{example}

\subsection{Optimizations}
\label{subsec:optimizations}
\btitle{In-Memory Edge Caching.}
As the edge set typically dominates the size of a graph \cite{DBLP:journals/tkdd/WangWWYY20}, especially for social networks and web graphs.
To extend the scale of graphs that our framework can process, it is crucial to reduce the memory footprint of edge data.
To achieve this, the edge data of our framework is not loaded into memory by default. Instead, it is loaded from external storage as needed and cached in memory when sufficient space is available.
In cases where memory is insufficient, we employ an edge caching system that uses a least-recently-used (LRU) policy to evict old edges from the memory cache and keep the most frequently accessed edges in memory.
As for graph updates, when they are deployed, they are temporarily stored in memory.
Once the neighbor edges are scheduled for eviction, the updated neighbor sets are written back to external storage.

\btitle{Critical Attributes.}
A vertex may have multiple custom attributes, some of which are accessed by other vertices in pull computation mode and must be synchronized to their mirrors, while others are used solely for local computation and are never accessed by other vertices.
For example, in \algpr, the attribute $\kw{r}$ is accessed by neighboring vertices, it is a critical attribute and must be synchronized to its mirrors. While in \algprDelta, the attribute \kw{dlt} is only used for local computation and never accessed by their neighbors.
To this end, we classify attributes into critical and regular categories.
An attribute is considered critical if it is accessed by other vertices, while regular attributes are used solely for local computation.
Only critical attributes are kept in mirrors and require synchronization when modified.
By statically analyzing the algorithm, we can identify the critical attributes in advance.
This optimization reduces unnecessary communication and memory overhead of mirrors for synchronizing and storing the non-critical attributes.

\btitle{Overlapping Communication with Computation.}
To improve hardware utilization and overall performance, we design strategies to overlap communication with computation.
During the computation phase, a worker may send data to other workers, such as attribute synchronization or message passing.
Before computation ends, some data may already be ready for transmission, while others are still being prepared.
To mitigate communication latency, we use two threads: one for computation and one for communication.
The computation thread is responsible for executing vertex programs and preparing data to be sent, while the communication thread is responsible for sending and receiving messages.
Once the computation thread finishes preparing a batch of data, it adds the batch to a pending message queue and notifies the communication thread.
The communication thread then sends the messages in the queue and monitors and receives incoming messages from other workers.

\subsection{Fault Tolerance}
Since dynamic graph algorithms are invoked automatically by the framework to process graph updates, a simple yet effective fault-tolerance mechanism is to store vertex states on disk after each algorithm invocation. If an algorithm fails, it can be restarted from the last saved state using the same batch of updates.

For algorithms that may take a long time to complete, a finer-grained fault-tolerance mechanism is necessary. As our framework support user-defined dynamic and complex control flows (e.g., loops and conditionals) and local variables, which are not managed by the framework, it is challenging to provide a fully transparent fault-tolerance solution.
A more practical approach is to provide an API that stores intermediate computation states, including vertex states, in-transit messages, and vertex subsets, in persistent storage. Users can invoke this API to create snapshots of the system state between supersteps.
Recovery of local control flows and variables should be handled by users. For example, they can explicitly store local variables and control flow information in the snapshots. During recovery, these variables can be restored from the snapshots, and the control flow can be resumed from the last snapshot. The implementation of this checkpoint mechanism is planned as future work for production deployment.
\section{Applications}
\label{sec:app}
This section presents several applications of our framework in addition to the PageRank example in \refsec{framework}.

\begin{algorithm}[t]
\caption{$\algsssp$}
\KwIn{Distance $dis$ and predecessor $pre$ of each vertex; graph update $\Delta G$}
\label{alg:sssp}
\tcc{$\apivertex(\kwint~\kw{dis},\kwint~\kw{pre})$}

$A \gets \apiloadupdate(\Delta)$; $A.\apideployupdate()$\;
$A \gets A.\apifilter(dis \geq 0)$\;

$D \gets A.\apiexplore(\apiforins\!: \mathcal{E}, \kw{minD}).\apicompute(\mathcal{C})$\;

    $B \gets A.\apiexplore(\apifordel\!: \mathcal{E'}); C \gets \emptyset$\;
    \While{$|B|>0$}{
        $B \gets B.\apicompute(\mathcal{C'}).\apifilter(pre=-1); C \gets C \cup B$\;
        $B \gets B.\apiexplore(\apifornb\!: \mathcal{E'})$\;
    }
    $C \gets C.\apicompute(\mathcal{C''})\cup D$\;
    \While{$|C|>0$}{
        $C \gets C.\apiexplore(\apifornb\!: \mathcal{E}, \kw{minD}).\apicompute(\mathcal{C})$\;
    }
\vspace{0.5em}
\stat{$\mathcal{E}$}{\tcp{Explore Possible Descendants}
    \If{$dis+1<nb.dis \vee nb.dis=-1$}{
        $\apisend(nb.id, Msg(dis+1,id))$\;
    }
}
\vspace{0.5em}
\stat{$\mathcal{E'}$}{
    \tcp{Explore Current Descendants}
    \lIf{$id=nb.pre$}{$\apiactivate(nb.id)$}
}
\vspace{0.5em}
\fn{$\kw{minD}(msgs)$}{\Return{$\arg\min_{m \in msgs} m.dis$}}
\vspace{0.5em}
\stat{$\mathcal{C}$}{$m \gets \arg\min_{m \in msgs} m.dis; dis\gets m.dis;pre \gets m.pre$\;}
\vspace{0.5em}
\stat{$\mathcal{C'}$}{
    \tcp{Update Predecessor}
    $pre \gets -1$\;
    $\apifornb\!:$\\
        \Indp \lIf{$dis=nb.dis+1 \wedge nb.pre \neq -1$}{$pre \gets nb.id$} \Indm
}
\vspace{0.5em}
\stat{$\mathcal{C''}$}{
    \tcp{Update Distance \& Predecessor}
    $dis \gets -1$\;
    $\apifornb\!:$\\
        \Indp \If{$(dis>nb.dis+1 \vee dis=-1) \wedge nb.dis \geq 0$}{$dis\gets nb.dis+1;pre \gets nb.id$\;} \Indm
}

\end{algorithm}

\subsection{Single-Source Shortest Path (SSSP)}
Given an unweighted graph $G(V,E)$ and a source vertex $s \in V$, the single-source shortest path problem computes the shortest paths from $s$ to all reachable vertices \cite{DBLP:conf/spaa/LeisersonS10,DBLP:conf/spaa/ShunBFGKST12}.
The pseudocode for our implementation is shown in \Cref{alg:sssp}.
For edge insertions, we identify affected vertices reachable from the source and send messages to newly inserted neighbors using a combiner (lines 3, 11--13, 16--17). Neighbors update their distance and predecessor if a shorter path is found (lines 18--19). Whenever a vertex's distance or predecessor changes, the update propagates to its neighbors (lines 8--10) until no further changes occur.
For edge deletions, we identify affected vertices reachable from the source and send messages to deleted successors (lines 4, 14--15). If a successor loses its predecessor, we search for an alternative predecessor with the same distance (line 6). If no alternative exists, the vertex is added to a separate set for further processing (line 6). For vertices with lost predecessors and no alternative, the process propagates to successors until none remain (lines 5--7). Each such vertex then attempts to find a new predecessor and update its distance (lines 8, 24--28). Whenever a vertex's distance or predecessor changes, the update propagates to its neighbors (lines 9--10) until convergence.

\begin{algorithm}[t]
\caption{$\algcc$}
\KwIn{Component ID $\kw{cid}$ and predecessor $\kw{pre}$ of each vertex; graph update $\Delta G$}
\label{alg:cc}
\tcc{$\apivertex(\kwlong~\kwlong~\kw{cid},\kwint~\kw{pre})$}

$A \gets \apiloadupdate(\Delta G)$\;
$A.\apideployupdate()$\;

    \tcc{Handle edge insertions}
    $B \gets A.\apiexplore(\apiforins\!: \mathcal{E}).\apicompute(\mathcal{C})$\;
    \While{$|B|>0$}{
    	$B \gets B.\apiexplore(\apifornb\!: \mathcal{E}).\apicompute(\mathcal{C})$\;
    }

    \tcc{Handle edge deletions}
    $B \gets A.\apiexplore(\apifordel\!: \mathcal{E'}); C \gets \emptyset$\;
    \While{$|B|>0$}{
        $B.\apicompute(\kw{cid} \gets deg*|V|+id;pre \gets -1;)$\;
        $ C \gets C \cup B$\;
        $B \gets B.\apiexplore(\apifornb\!: \mathcal{E'})$\;
    }
    \While{$|C|>0$}{
        $C \gets C.\apicompute(\mathcal{C}).\apiexplore(\apifornb\!: \mathcal{E})$\;
    }

\vspace{0.5em}
\stat{$\mathcal{C}$}{
    $\apifornb\!:$\\
        \Indp \lIf{$\kw{cid} < nb.\kw{cid}$}{$\kw{cid}\gets nb.\kw{cid},pre \gets nb.id$} \Indm
}
\vspace{0.5em}
\stat{$\mathcal{E}$}{
    \lIf{$\kw{cid} > nb.\kw{cid}$}{$\apiactivate(nb.id)$}
}
\vspace{0.5em}
\stat{$\mathcal{E'}$}{
    \lIf{$id = nb.pre$}{$\apiactivate(nb.id)$}
}
\end{algorithm}

\subsection{Connected Components (CC)}

Given an undirected graph $G(V,E)$, the connected components are vertex sets in which every pair of vertices is mutually reachable, and no vertex is reachable across different sets \cite{DBLP:conf/sigmod/QinYCCZL14}.
The pseudocode for our implementation is shown in \Cref{alg:cc}.
For edge insertions, we identify affected vertices and send messages to newly inserted neighbors with a smaller CC ID \kw{cid} (lines 3, 15--16). Each such neighbor sets its \kw{cid} to the maximum \kw{cid} among its neighbors (lines 3, 12--14). If a vertex's \kw{cid} or predecessor updates, the process propagates to its neighbors until no further updates occur (lines 4--5).
For edge deletions, we identify affected vertices and send messages to deleted successors (lines 6, 18--19). If a successor loses its predecessor, it resets \kw{cid} to its own ID, clears the predecessor (line~8), and enters a separate set for further processing (line~9). For vertices that lost predecessors, propagation continues to their successors until no successors remain (line~10).
For each vertex in the separate set, we recompute \kw{cid} from neighbors; if \kw{cid} changes, we propagate to neighbors with smaller \kw{cid} until convergence (lines~11--12).

\subsection{Maximal Independent Set (MIS)}
Given an undirected graph $G(V,E)$, a maximal independent set is a set of vertices such that no two vertices are adjacent (independence), and no other vertex can be added without violating this property \cite{DBLP:conf/stoc/Luby85, DBLP:journals/siamcomp/Luby86, DBLP:journals/jal/AlonBI86}.
The pseudocode for our implementation is shown in \Cref{alg:mis}.
The flag \kw{s} indicates whether a vertex is in the MIS.
We first identify the affected vertices and their neighbors (line 3). For each vertex (line 5), we set its MIS flag \kw{s} to true (line 7). Then, for each neighbor, if the neighbor's flag is true and it has a lower ranking than the vertex, we set \kw{s} to false and stop processing its neighbors (lines 8--9).
The ranking function is defined as:
{\small
\[
\mathcal{R}(a) > \mathcal{R}(b)
\iff a.deg > b.deg \ \vee\
(a.deg = b.deg \wedge a.id > b.id).
\]
}
Next, for each vertex whose \kw{s} flag changes (line 5), we activate its neighbors if they have a higher ranking and share the same \kw{s} value (lines 10--12).
This update-propagation process repeats until no more updates occur (lines 4--5).

\begin{algorithm}[t]
\caption{$\algmis$}
\KwIn{Flag $\kw{s}$ of each vertex; graph update $\Delta G$}
\label{alg:mis}
\tcc{$\apivertex(\kwbool~\kw{s})$}

$A \gets \apiloadupdate(\Delta G)$\;
$A.\apideployupdate()$\;

    $B \gets A \cup A.\apiexplore(\apifornb\!: \apiactivate(nb.id))$\;
    \While{$|B|>0$}{
    	$B \gets B.\apicompute(\mathcal{C}).\apiexplore(\mathcal{E})$\;
    }
\vspace{0.5em}
\stat{$\mathcal{C}$}{
    $s \gets \kw{True}$\;
    $\apifornb\!:$\\
        \Indp \lIf{$\mathcal{R}(v) > \mathcal{R}(nb) \wedge nb.s$}{$s \gets \kw{False};\kw{Break}$} \Indm
}
\vspace{0.5em}
\stat{$\mathcal{E}$}{
    $\apifornb\!:$\\
        \Indp \lIf{$\mathcal{R}(v) < \mathcal{R}(nb) \wedge s = nb.s$}{$\apiactivate(nb.id)$} \Indm
}
\end{algorithm}

\begin{algorithm}[t]
\caption{$\algcolor$}
\KwIn{Previous degree $\kw{od}$, previous color $\kw{oc}$, and color $\kw{c}$ of each vertex; graph update $\Delta G$}
\label{alg:color}
\tcc{$\apivertex(\kwint~\kw{od},\kwint~\kw{oc},\kwint~\kw{c})$}

$A \gets \apiloadupdate(\Delta)$\;
$A.\apicompute(\kw{od} \gets deg)$\;
$A.\apideployupdate()$\;

    $B \gets A \cup A.\apiexplore(\mathcal{E})$\;
    \While{$|B|>0$}{
        $B \gets B.\apicompute(\mathcal{C}).\apiexplore(\mathcal{E'})$\;
    }
\vspace{0.5em}
\stat{$\mathcal{C}(v)$}{
    $U[0..|V|] \gets \kw{False}$\;
    $\apifornb\!:$\\
        \Indp \lIf{$\mathcal{R}(v) < \mathcal{R}(nb)$}{$U[nb.\kw{c}] \gets \kw{True}$} \Indm
    \For{$i \gets 0$ \KwTo $|V|$}{
        \lIf{$\neg U[i]$}{$\kw{oc} \gets \kw{c};\kw{c} \gets i;\kw{Break}$}
    }
}
\vspace{0.5em}
\stat{$\mathcal{E}$}{
    $\apifornb\!:$\\
        \Indp \lIf{$(\mathcal{R}(v) < \mathcal{R}(nb) \wedge \mathcal{R'}(v) > \mathcal{R}(nb) \wedge \kw{c} < nb.\kw{c})$ $\vee$ $(\mathcal{R}(v) > \mathcal{R}(nb) \wedge \mathcal{R'}(v) < \mathcal{R}(nb) \wedge \kw{c} = nb.\kw{c})$}{$\apiactivate(nb.id)$} \Indm
}
\vspace{0.5em}
\stat{$\mathcal{E'}$}{
    $\apifornb\!:$\\
        \Indp \lIf{$\mathcal{R}(v) > \mathcal{R}(nb) \wedge (\kw{c} = nb.\kw{c} \vee \kw{oc} < nb.\kw{c})$}{$\apiactivate(nb.id)$} \Indm
}
\end{algorithm}

\subsection{Graph Coloring (GC)}

Given an undirected graph $G(V,E)$, graph coloring assigns colors to vertices so that no two adjacent vertices share the same color. Among existing graph coloring methods, the Global algorithm \cite{DBLP:journals/pvldb/YuanQLCZ17} is widely adopted for its simplicity and effectiveness. It processes vertices in non-decreasing degree order, breaking ties by vertex ID, and assigns each the smallest color not used by its colored neighbors. \Cref{alg:color} implements a dynamic Global algorithm for graph color updates.
We first record the previous degree \kw{od} of each affected vertex before deploying the update (line 2). We then collect these vertices and their neighbors that satisfy one of the following conditions into a set (line 4):
(1) the neighbor has a higher color number than the vertex, a higher ranking than the vertex after the update, and a lower ranking than the vertex before the update; or
(2) the neighbor has the same color number as the vertex, a lower ranking than the vertex after the update, and a higher ranking than the vertex before the update (lines 13--15).
The ranking functions are defined as:
{\small
\[
\mathcal{R}(a) > \mathcal{R}(b)
\iff a.deg > b.deg \ \vee\
(a.deg = b.deg \wedge a.id > b.id),
\]
\[
\mathcal{R'}(a) > \mathcal{R}(b)
\iff a.\kw{od} > b.deg \ \vee\
(a.\kw{od} = b.deg \wedge a.id > b.id),
\]
\[
\mathcal{R'}(a) < \mathcal{R}(b)
\iff a.\kw{od} < b.deg \ \vee\
(a.\kw{od} = b.deg \wedge a.id > b.id).
\]
}
For each vertex in the set, we compute its color \kw{c} and propagate updates to its neighbors that meet the propagation condition, forming a new set. This process repeats until the set becomes empty (lines 5--6).
Color computation initializes a boolean array $U$ to track used colors, marks the colors of higher-ranked neighbors as used, records the vertex's previous color $oc$, and assigns it the smallest unused color (lines 7--12).
During propagation, if a neighbor has a lower ranking than the vertex, the same color number as the vertex, and a larger color number than the vertex before the update, it is added to the propagation set (lines 16--18).
\section{Experiments}\label{sec:exp}

\begin{table}
	\caption{Dataset Statistics}
	\label{tab:dataset}
	\centering
	\setlength{\tabcolsep}{3pt}
    \begin{tabular}{lrrl}
        \toprule
		Graph                                                                & $|V|$      & $|E|$         & Graph Type \\
        \midrule
		US-Road (UR) \cite{DBLP:conf/www/Kunegis13}                          & 23,947,347 & 28,854,312    & Road \\
		LiveJournal (LJ) \cite{DBLP:journals/tist/LeskovecS16}               & 4,847,571  & 42,851,237    & Social \\
		Euro-Road (ER) \cite{DBLP:conf/aaai/RossiA15}                        & 50,912,018 & 54,054,660    & Road \\
		Wikipedia (WK) \cite{DBLP:conf/www/Kunegis13}                        & 13,593,032 & 334,591,525   & Web \\
		UK-2005 (UK) \cite{DBLP:conf/www/BoldiV04, DBLP:conf/www/BoldiRSV11} & 39,459,925 & 783,027,125   & Web \\
		IT-2004 (IT) \cite{DBLP:conf/www/BoldiV04, DBLP:conf/www/BoldiRSV11} & 41,291,594 & 1,027,474,947 & Web \\
		Twitter (TW) \cite{DBLP:conf/www/Kunegis13}                          & 41,652,230 & 1,202,513,046 & Social \\
		Friendster (FS) \cite{DBLP:conf/www/Kunegis13}                       & 65,608,366 & 1,806,067,135 & Social \\
        \bottomrule
	\end{tabular}
\vspace{-1em}
\end{table}

\subsection{Experimental Setup}
\textbf{Platforms.}
Distributed experiments were conducted on a cluster of eight cloud instances, each with 16~vCPUs and 128~GB of RAM. We ran one MPI worker process per vCPU.

\btitle{Datasets and Updates.}
The statistics of the eight real-world graphs are given in \reftab{dataset}. We remove self-loops and duplicate undirected edges, order the edges using a fixed seed, and use the first half to construct the initial graph. We draw deletions from the initial graph and insertions from the other half. Both systems receive the same initial graph and ordered updates. By default, we apply a single batch of 2~million updates, with 1~million insertions and 1~million deletions. To vary the insertion ratio, we instead divide a stream of 200,000 updates into 20 consecutive batches of 10,000. The specified ratio applies to the whole stream.

\btitle{Applications and Baseline.}
We evaluate PageRank (PR), Single-Source Shortest Path (SSSP), Connected Components (CC), Maximal Independent Set (MIS), Graph Coloring (GC), and Core Decomposition (CD). Our distributed baseline is Ingress~\cite{DBLP:journals/pvldb/GongTYYZGYYZ21,DBLP:journals/vldb/GongTYWYZYGFYZ24}. We compare the two systems on PR, SSSP, and CC, and evaluate MIS, GC, and CD on our system alone. For SSSP, all edges have unit weight. For each graph, we sample non-isolated source candidates with a fixed seed and use the same selected vertex in both systems. For PR, both systems use a damping factor of 0.85.
Our system propagates changes above $10^{-12}$, and Ingress stops when the change at every vertex falls below this threshold. The two implementations also handle dangling vertices differently.

We report update time after the initial graph computation. For our system, timing begins when the workers are ready to read a batch. It covers reading and distributing the changed edges, applying them to the resident graph, updating the vertex state, and synchronizing and clearing the batch. Ingress keeps its workers active between batches. Its timer begins once the updated snapshot is available to all workers. It includes loading the changed edges, repairing dependencies, reloading the snapshot, and completing the incremental computation and synchronization. The producer constructs the snapshot and the adapter transfers and verifies it before timing begins. For each batch, we take the longest update time among workers and sum these times over all batches. To summarize speedup, we divide the Ingress update time by ours for each graph with matching resource settings, then take the geometric mean.

\subsection{Distributed Performance}

\begin{figure}[!htbp]
    \centering
    \includegraphics[width=\linewidth]{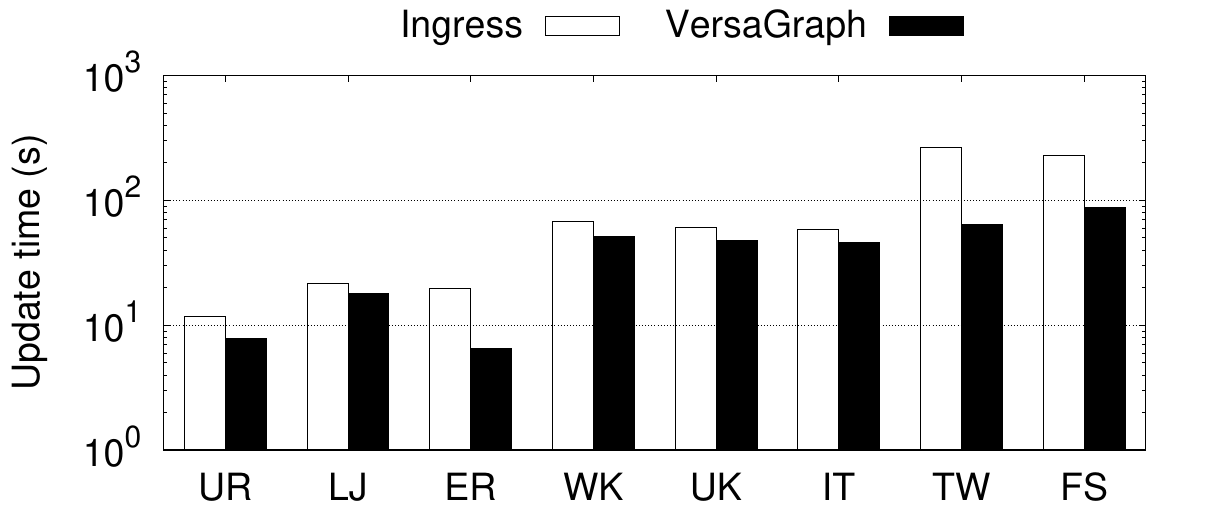}
    \caption{Update times for PR.}
    \label{fig:disBatchSize_PR}
    \vspace{-0.5em}
\end{figure}

\begin{figure}[!htbp]
    \centering
    \includegraphics[width=\linewidth]{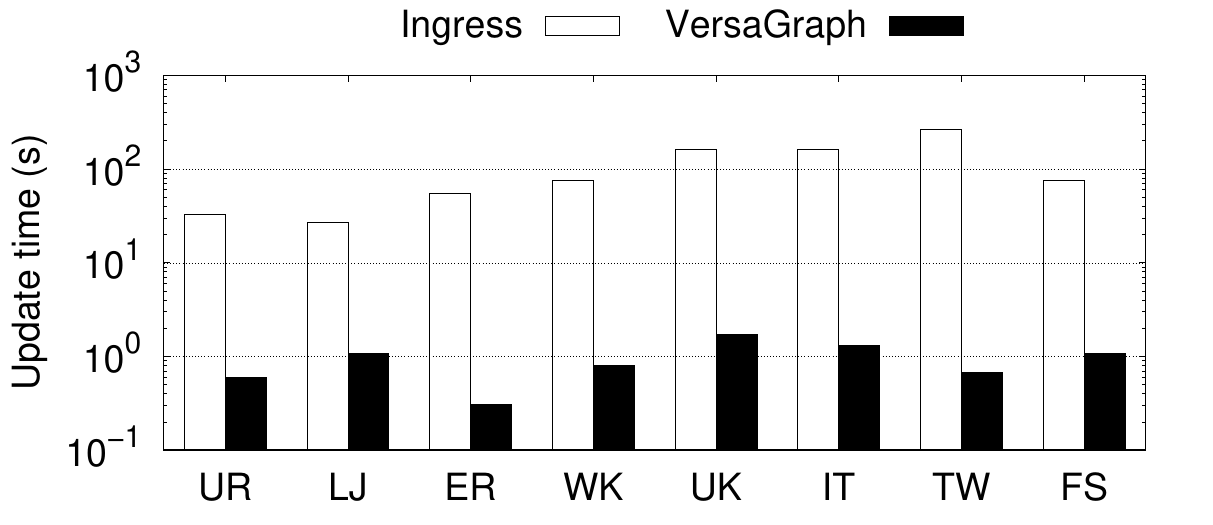}
    \caption{Update times for SSSP.}
    \label{fig:disBatchSize_BFS}
    \vspace{-0.5em}
\end{figure}

\begin{figure}[!htbp]
    \centering
    \includegraphics[width=\linewidth]{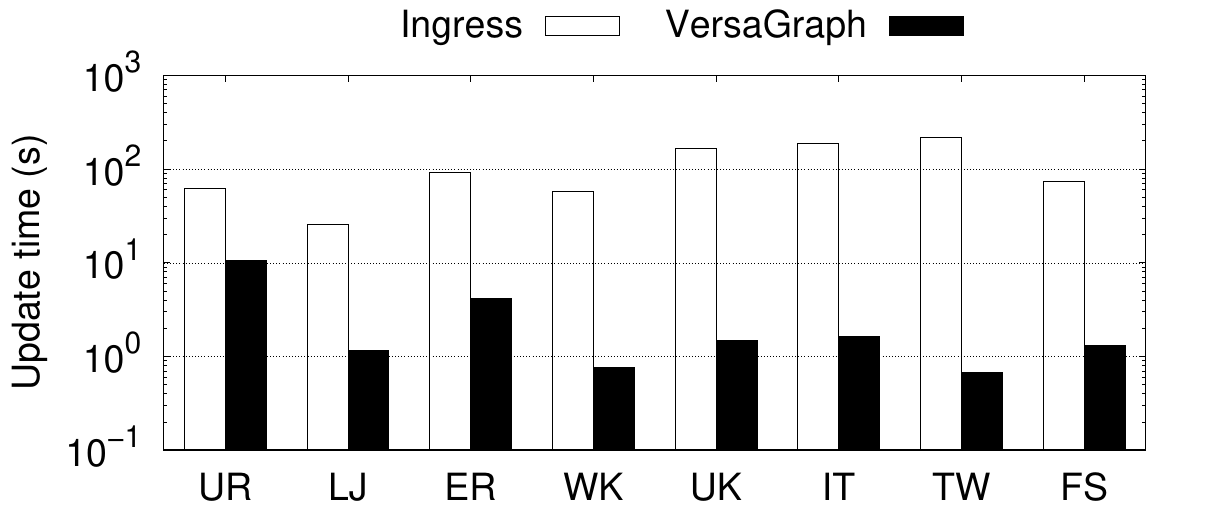}
    \caption{Update times for CC.}
    \label{fig:disBatchSize_CC}
    \vspace{-0.5em}
\end{figure}

We first evaluate PR, SSSP, and CC across all eight datasets using a batch size of 2M and 128 workers.
\Cref{fig:disBatchSize_PR} shows the update times for PR. Our system records a shorter time on each dataset. On IT, for example, our system takes 46.36~s and Ingress takes 58.68~s with matched resources.

\Cref{fig:disBatchSize_BFS} shows the update times for SSSP. On UR, LJ, WK, and UK, both systems have the same RAM per instance and use the same worker placement and thread affinity. Across these four graphs, our system achieves a geometric mean speedup of $59.17\times$. On UR, the update takes 0.60~s on our system and 32.59~s on Ingress. For Ingress, the longest dependency-repair preparation phase takes 22.66~s and the longest solver phase takes 0.27~s.

\Cref{fig:disBatchSize_CC} shows the corresponding results for CC. Our system achieves a geometric mean speedup of $30.05\times$ on the five graphs with matched settings (UR, LJ, ER, WK, and UK). On FS, the RAM available per instance differs between the two systems. Our system takes 1.30~s, and Ingress takes 73.81~s with the compact dependency tree.

\begin{figure}[!htbp]
    \centering
    \begin{subfigure}[t]{0.74\linewidth}
        \centering
        \includegraphics[width=\linewidth]{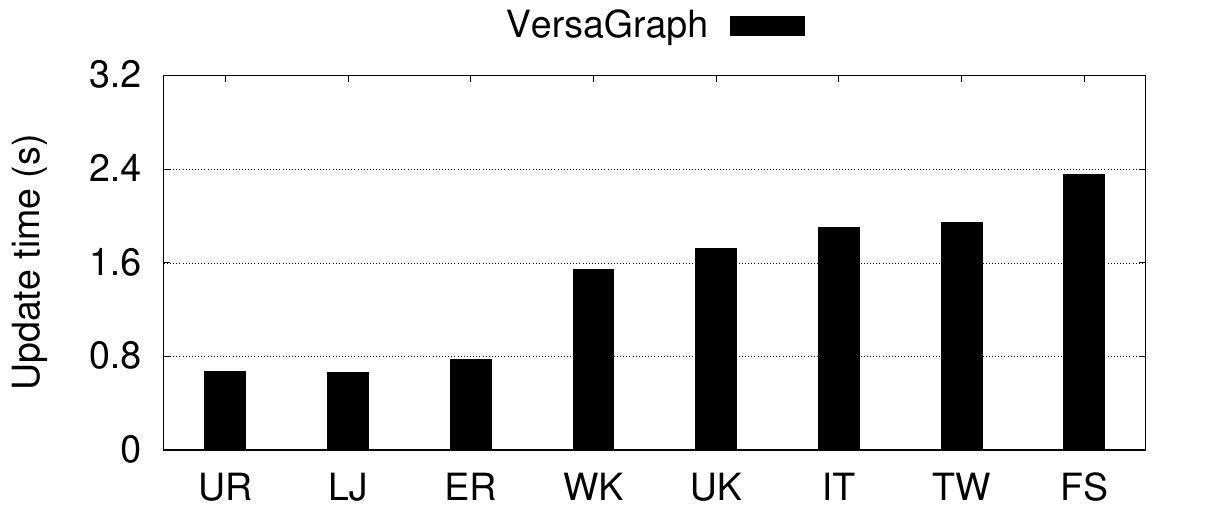}
        \caption{MIS}
    \end{subfigure}\par
    \begin{subfigure}[t]{0.74\linewidth}
        \centering
        \includegraphics[width=\linewidth]{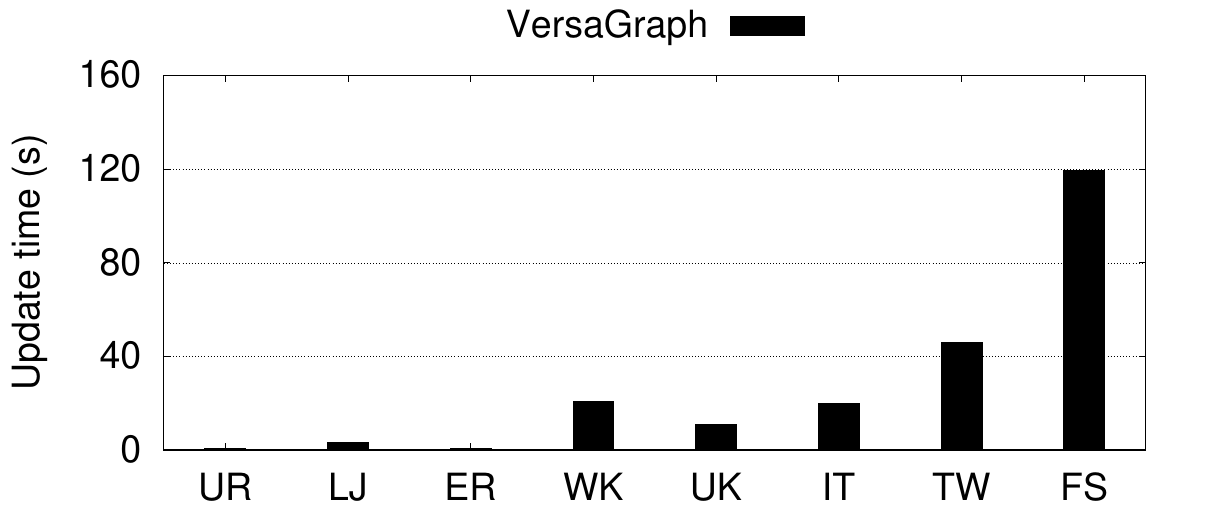}
        \caption{GC}
    \end{subfigure}\par
    \begin{subfigure}[t]{0.74\linewidth}
        \centering
        \includegraphics[width=\linewidth]{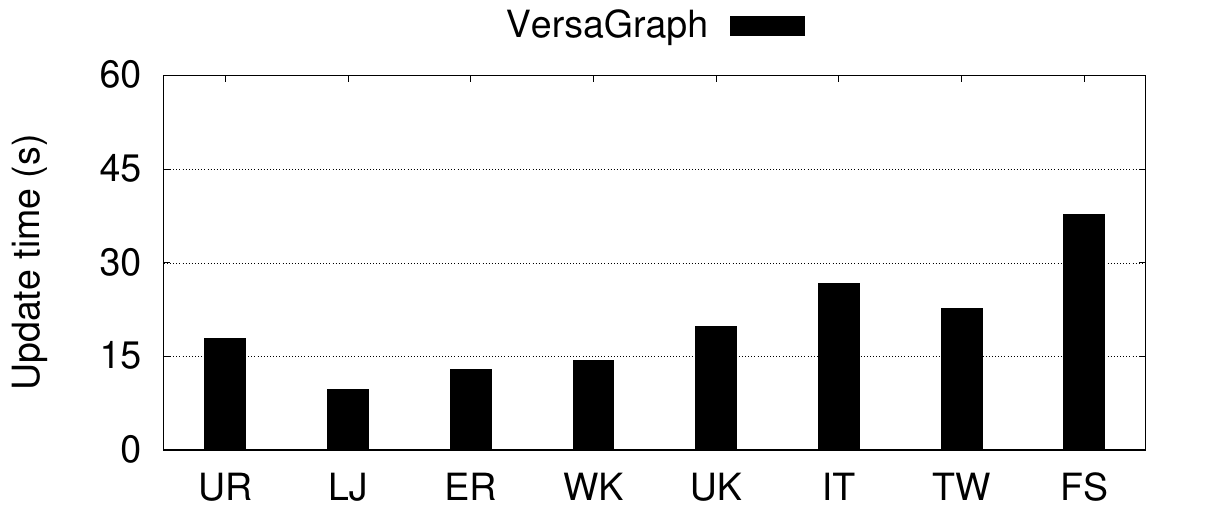}
        \caption{CD}
    \end{subfigure}\par
    \caption{Update times for MIS, GC, and CD on our system.}
    \label{fig:disAdditional}
    \vspace{-0.5em}
\end{figure}

\Cref{fig:disAdditional} shows the results for the other three applications. MIS takes between 0.66 and 2.36~s across the eight graphs. GC varies more widely, from 0.70~s on UR to 119.60~s on FS. CD takes 9.72~s on LJ and 37.79~s on FS, the lowest and highest times observed for this application.

\begin{figure}[!htbp]
    \centering
    \includegraphics[width=0.59\linewidth,trim=308bp 0bp 275bp 0bp,clip]{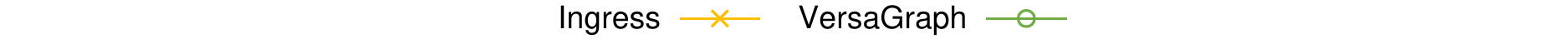}\par
    \begin{subfigure}[t]{0.32\linewidth}
        \centering
        \includegraphics[width=\linewidth]{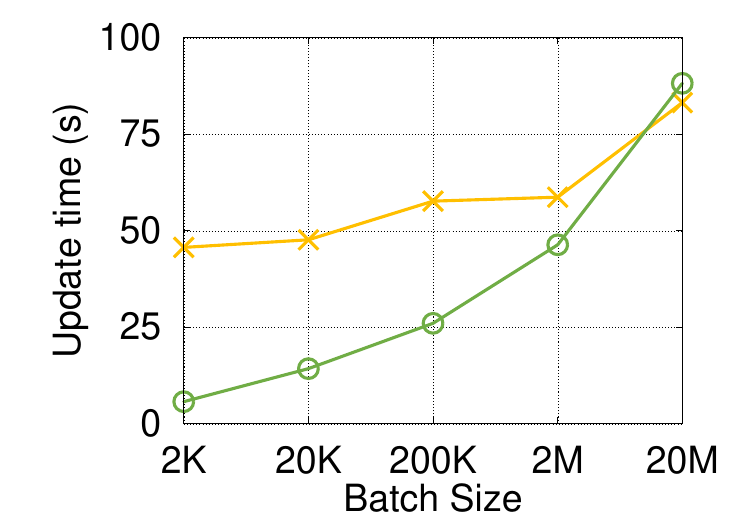}
        \caption{PR (IT)}
    \end{subfigure}\hfill
    \begin{subfigure}[t]{0.32\linewidth}
        \centering
        \includegraphics[width=\linewidth]{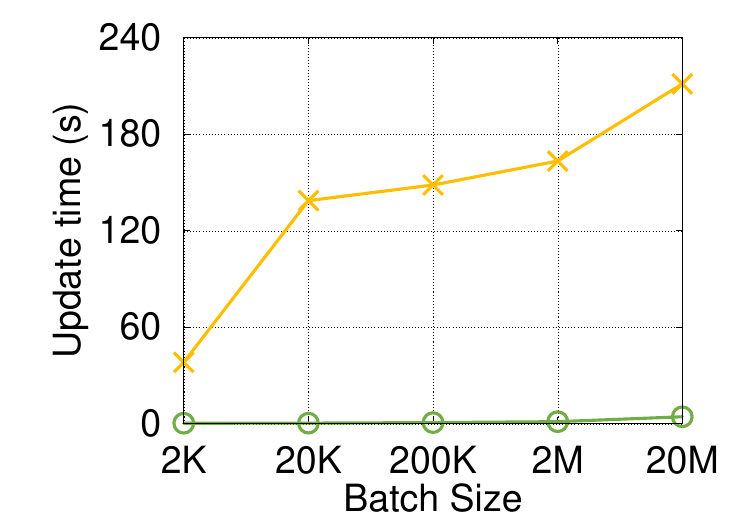}
        \caption{SSSP (IT)}
    \end{subfigure}\hfill
    \begin{subfigure}[t]{0.32\linewidth}
        \centering
        \includegraphics[width=\linewidth]{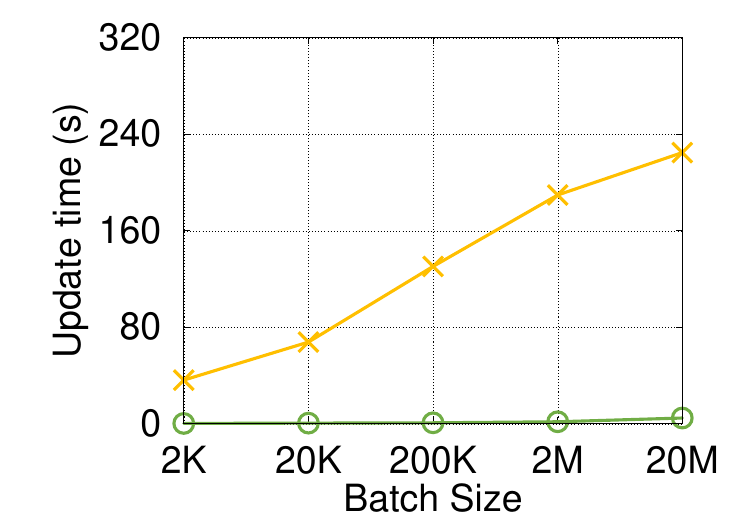}
        \caption{CC (IT)}
    \end{subfigure}\par
    \begin{subfigure}[t]{0.32\linewidth}
        \centering
        \includegraphics[width=\linewidth]{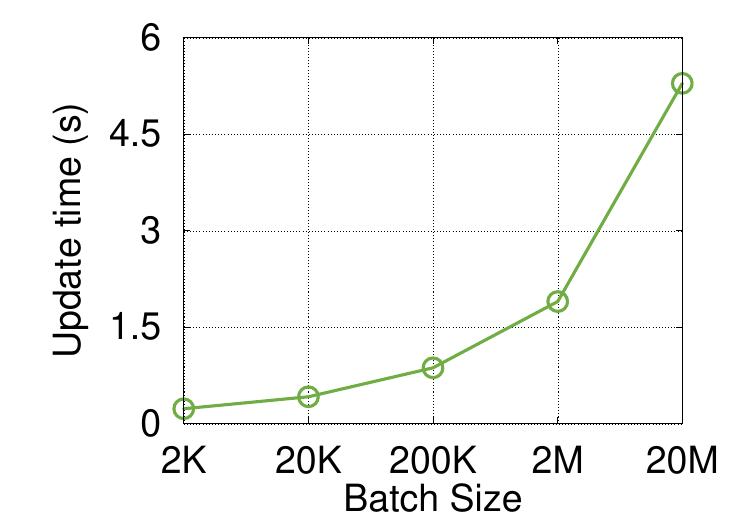}
        \caption{MIS (IT)}
    \end{subfigure}\hfill
    \begin{subfigure}[t]{0.32\linewidth}
        \centering
        \includegraphics[width=\linewidth]{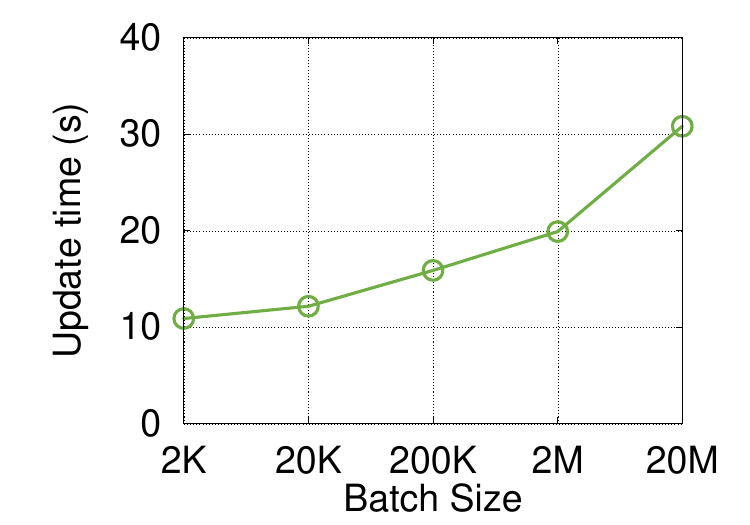}
        \caption{GC (IT)}
    \end{subfigure}\hfill
    \begin{subfigure}[t]{0.32\linewidth}
        \centering
        \includegraphics[width=\linewidth]{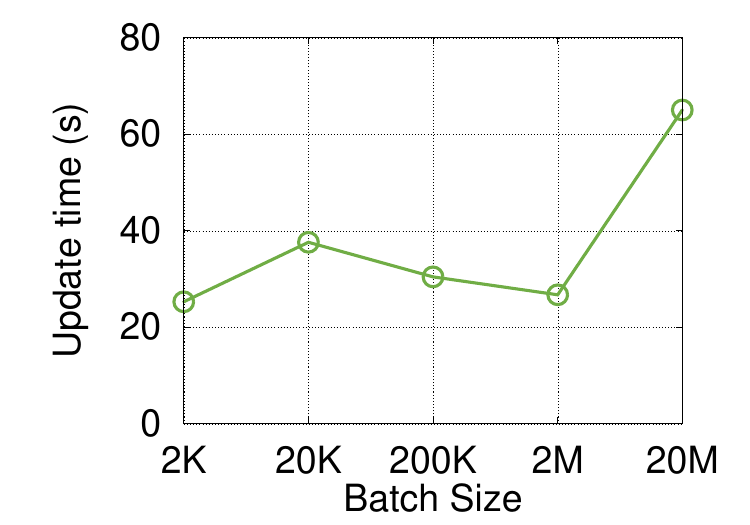}
        \caption{CD (IT)}
    \end{subfigure}\par
    \begin{subfigure}[t]{0.32\linewidth}
        \centering
        \includegraphics[width=\linewidth]{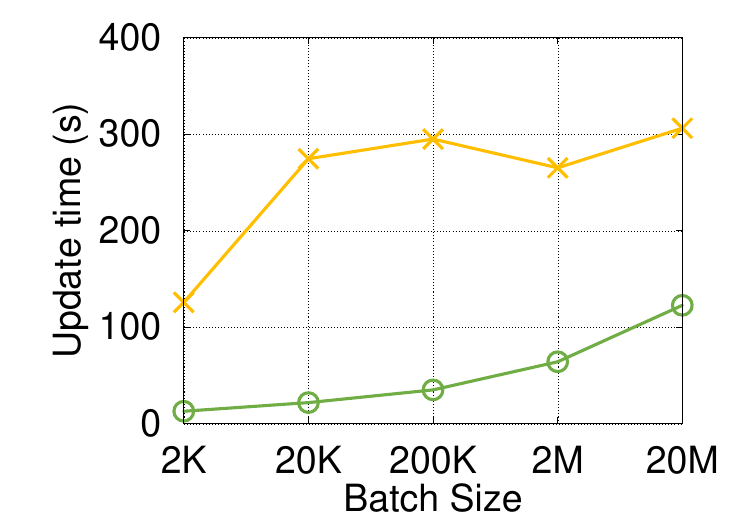}
        \caption{PR (TW)}
    \end{subfigure}\hfill
    \begin{subfigure}[t]{0.32\linewidth}
        \centering
        \includegraphics[width=\linewidth]{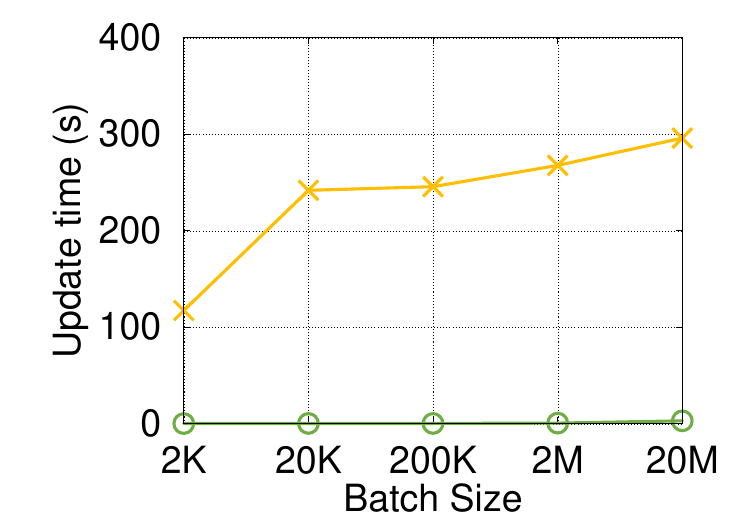}
        \caption{SSSP (TW)}
    \end{subfigure}\hfill
    \begin{subfigure}[t]{0.32\linewidth}
        \centering
        \includegraphics[width=\linewidth]{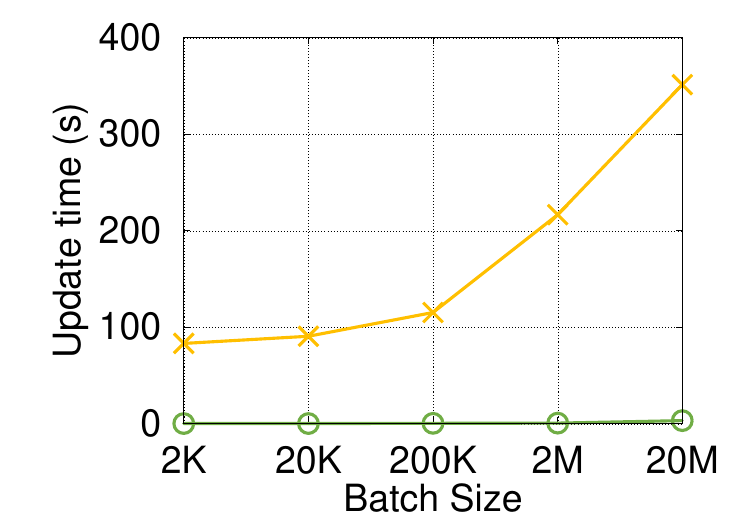}
        \caption{CC (TW)}
    \end{subfigure}\par
    \begin{subfigure}[t]{0.32\linewidth}
        \centering
        \includegraphics[width=\linewidth]{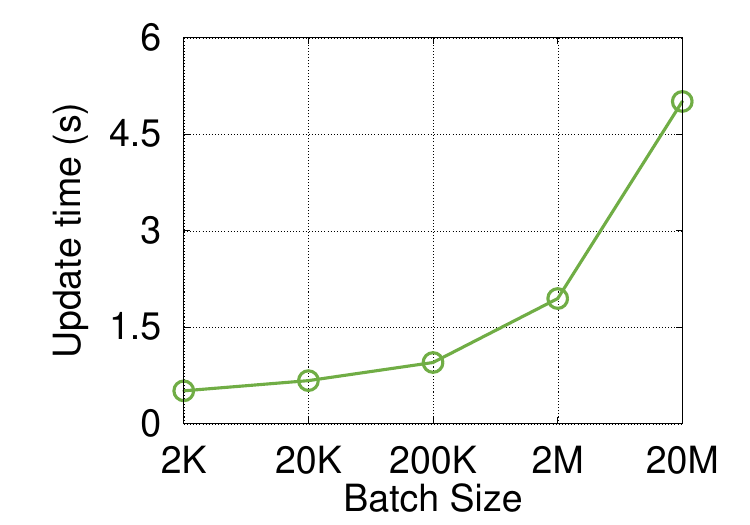}
        \caption{MIS (TW)}
    \end{subfigure}\hfill
    \begin{subfigure}[t]{0.32\linewidth}
        \centering
        \includegraphics[width=\linewidth]{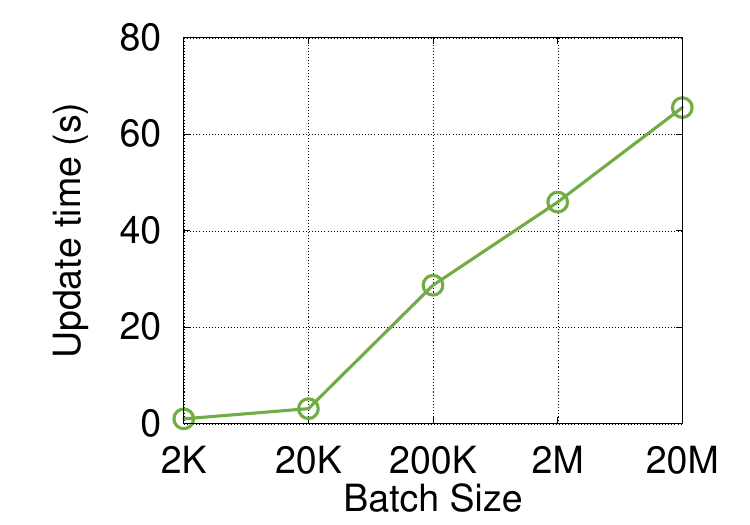}
        \caption{GC (TW)}
    \end{subfigure}\hfill
    \begin{subfigure}[t]{0.32\linewidth}
        \centering
        \includegraphics[width=\linewidth]{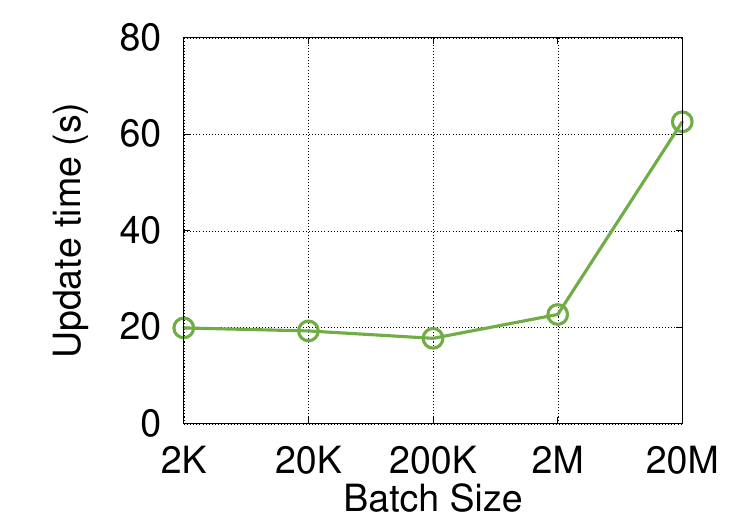}
        \caption{CD (TW)}
    \end{subfigure}\par
    \caption{Update times by varying batch size.}
    \label{fig:disSize}
    \vspace{-0.5em}
\end{figure}

We next evaluate all six algorithms on IT and TW using 128 workers, varying the batch size across 2K, 20K, 200K, 2M, and 20M updates, as shown in \Cref{fig:disSize}. For PR on IT, our system is faster at the four smaller batch sizes. At 20M updates, Ingress records the shorter update time, taking 83.18~s compared with 88.17~s on our system. For SSSP, our system takes 0.30--4.34~s on IT and 0.23--3.05~s on TW. Even at 2K updates on IT, Ingress takes 38.18~s, including the reload of the updated graph. For GC on TW, increasing the batch size from 2K to 20M raises the time on our system from 1.02 to 65.51~s. CD takes 19.86~s at 2K, falls to 17.70~s at 200K, and reaches 62.59~s at 20M.

\begin{figure}[!htbp]
    \centering
    \includegraphics[width=0.59\linewidth,trim=308bp 0bp 275bp 0bp,clip]{figures/legend.pdf}\par
    \begin{subfigure}[t]{0.48\linewidth}
        \centering
        \includegraphics[width=\linewidth]{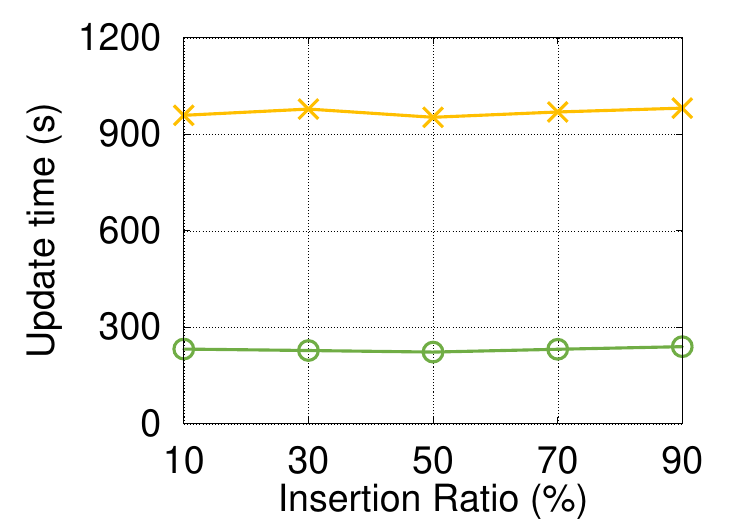}
        \caption{PR (IT)}
    \end{subfigure}\hfill
    \begin{subfigure}[t]{0.48\linewidth}
        \centering
        \includegraphics[width=\linewidth]{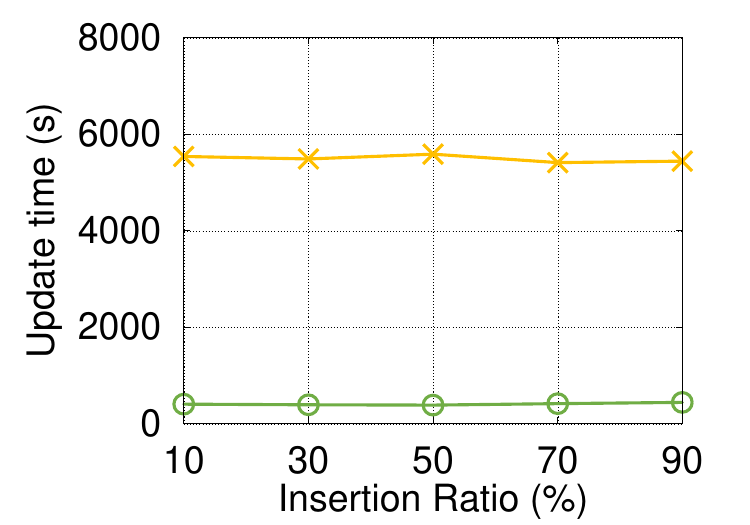}
        \caption{PR (TW)}
    \end{subfigure}\par
    \begin{subfigure}[t]{0.48\linewidth}
        \centering
        \includegraphics[width=\linewidth]{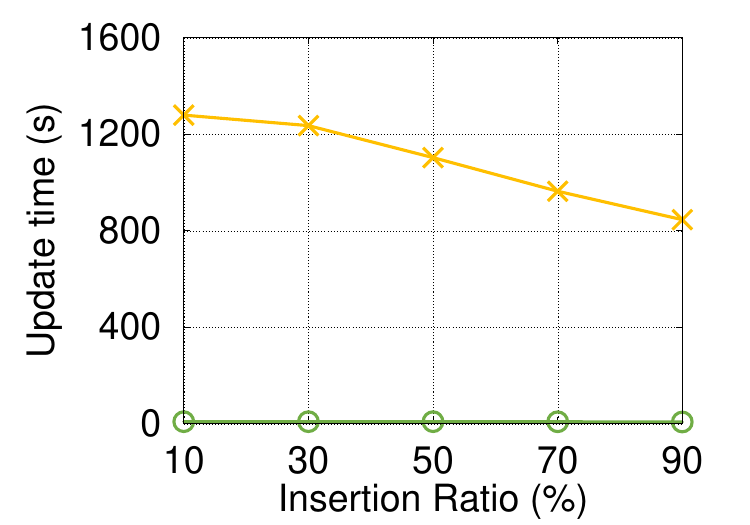}
        \caption{SSSP (IT)}
    \end{subfigure}\hfill
    \begin{subfigure}[t]{0.48\linewidth}
        \centering
        \includegraphics[width=\linewidth]{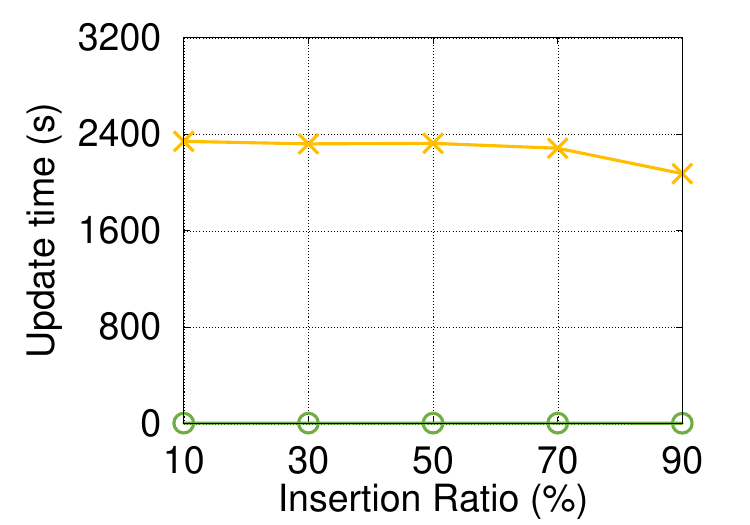}
        \caption{SSSP (TW)}
    \end{subfigure}\par
    \begin{subfigure}[t]{0.48\linewidth}
        \centering
        \includegraphics[width=\linewidth]{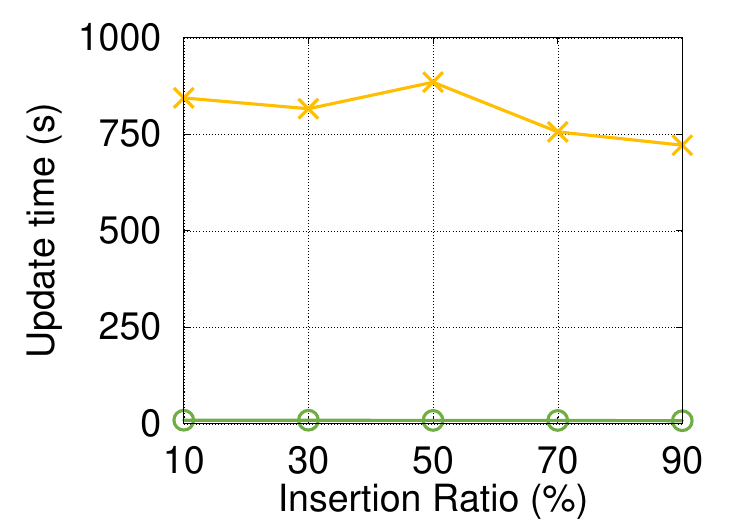}
        \caption{CC (IT)}
    \end{subfigure}\hfill
    \begin{subfigure}[t]{0.48\linewidth}
        \centering
        \includegraphics[width=\linewidth]{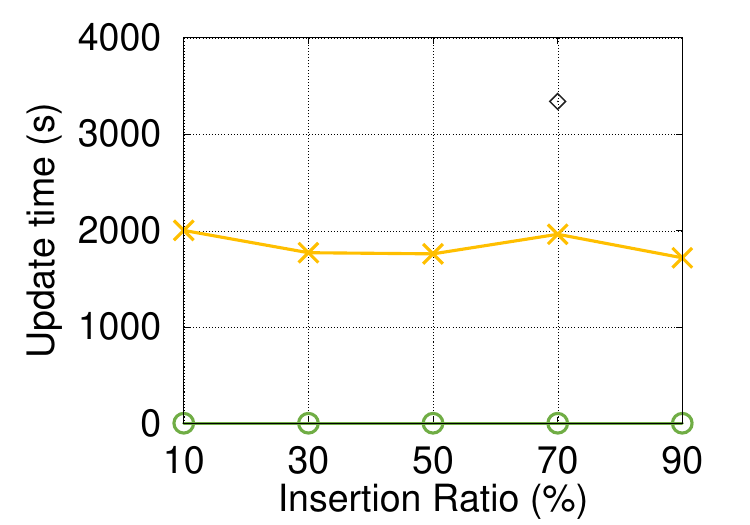}
        \caption{CC (TW)}
    \end{subfigure}\par
    \caption{Update times by varying the fraction of insertions over 20 batches. In (f), the line and the open diamond show the two measurements at 70\% insertions.}
    \label{fig:disRatioL8}
    \vspace{-0.5em}
\end{figure}

We vary the insertion ratio from 10\% to 90\% in steps of 20 percentage points, using 20 batches of 10K updates on IT and TW (\Cref{fig:disRatioL8}). For CC on IT, our system takes 7.73--9.17~s and Ingress takes 721.63--884.67~s with matched resources. For CC on TW at 70\% insertions, Ingress takes 1,963.47~s on the curve and 3,339.91~s at the open diamond. The last four batches account for 394.23~s and 1,946.79~s, respectively. In the shorter run, each of these batches takes 98.25--99.04~s, within the range of the first 16 batches, and total update time is 41.21\% lower. The neighboring 50\% and 90\% runs take 1,760.93 and 1,719.65~s, respectively.

\begin{figure}[!htbp]
    \centering
    \includegraphics[width=\linewidth]{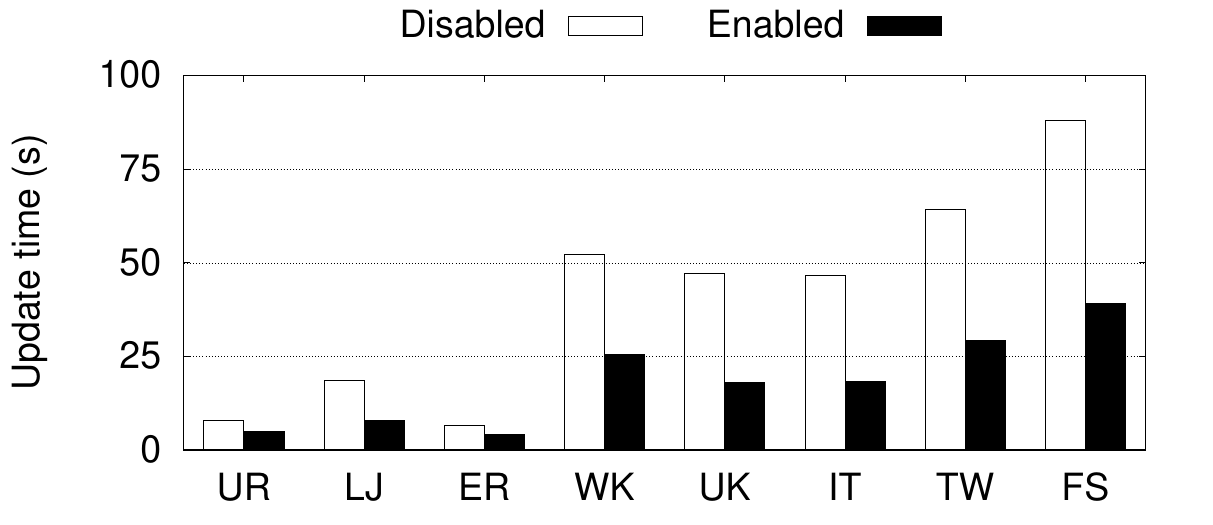}
    \caption{Update times for PR with and without the critical-attribute optimization.}
    \label{fig:disCritical_PR}
    \vspace{-0.5em}
\end{figure}

Finally, we evaluate the critical-attribute optimization introduced in \Cref{subsec:optimizations}. We run the same PR executable twice on each graph with 128 workers and a batch size of 2M. With the optimization disabled, workers synchronize all changed fields. With it enabled, they synchronize only the degree field. As shown in \Cref{fig:disCritical_PR}, enabling the optimization reduces the time on all eight graphs. The speedup ranges from $1.55\times$ on ER to $2.63\times$ on UK, with a geometric mean of $2.10\times$.
\section{Related Work}\label{sec:related}
A variety of systems have been proposed to support incremental and dynamic graph processing, each targeting different algorithmic properties and performance goals. These systems can be broadly categorized into four types: stateless systems, dependency-tracking systems, version-tracking systems, and automated systems.

\btitle{Stateless Systems.}
Stateless systems are designed for algorithms that converge to the same result regardless of initial conditions. These systems accelerate dynamic processing by using the previous output as the starting point for recomputation. Tornado~\cite{DBLP:conf/sigmod/ShiCST16} exemplifies this approach by leveraging the iterative nature of algorithms like PageRank to reduce recomputation time after updates.

\btitle{Dependency-Tracking Systems.}
For monotonic algorithms where values only move in one direction, dependency-tracking systems maintain information about which vertices contribute to the current result. When the graph changes, only affected vertices are updated based on tracked dependencies. Systems such as KickStarter~\cite{DBLP:conf/asplos/VoraGX17} and RisGraph~\cite{DBLP:conf/sigmod/FengML0ZHC21} fall into this category and focus on ensuring correctness with minimal recomputation.

\btitle{Version-Tracking Systems.}
Some algorithms require strict synchronization across rounds and depend on the full state of neighboring vertices. To support these cases, version-tracking systems retain intermediate results and selectively update values in response to graph changes. GraphBolt~\cite{DBLP:conf/eurosys/MariappanV19} and DZiG~\cite{DBLP:conf/eurosys/MariappanCV21} belong to this class, trading additional memory overhead for broader applicability.

\btitle{Automated Systems.}
Recent work has explored automated incrementalization by analyzing static algorithms and selecting suitable update strategies. Systems like Ingress~\cite{DBLP:journals/pvldb/GongTYYZGYYZ21, DBLP:journals/vldb/GongTYWYZYGFYZ24} and Layph~\cite{DBLP:conf/icde/YuGZY00TY0Z23} aim to reduce user burden by offering general-purpose support for dynamic updates based on formal analysis.

\btitle{Other Systems.}
Beyond these categories, several systems extend data-parallel frameworks to support graph updates. iGraph~\cite{DBLP:journals/fcsc/JuLYZ16} and IncGraph~\cite{DBLP:journals/tkde/TangHFY22} adapt Spark-based platforms for incremental graph processing. Differential Dataflow~\cite{DBLP:conf/cidr/McSherryMII13} offers general incremental computation but lacks graph-specific optimization. Systems like iTurboGraph~\cite{DBLP:conf/sigmod/KoLHLSSH21} focus on improving update handling through alternative execution models.

While these systems provide capabilities for dynamic graph processing, the majority of them are limited to single-machine environments and cannot scale to large graphs.
However, Ingress, the only distributed system designed for dynamic graphs, exhibits significant limitations, as discussed in \Cref{sec:introduction}.
\section{Conclusion}
\label{sec:conclusion}
In this paper, we presented a new system for dynamic graph processing that combines the flexibility of user-defined algorithm expression with high performance. It offers a simple programming model with minimal boilerplate and a small set of expressive core APIs supported by system-level optimizations for efficient execution. We evaluated the system on eight large-scale real-world datasets with six representative algorithms and found that it is flexible, expressive, and outperforms existing systems. This work bridges the gap between algorithmic customization and system efficiency and opens opportunities for future research on generalizable high-performance dynamic graph systems.

\end{document}